\documentclass[preprint]{elsarticle}

\usepackage{paralist}
\usepackage{amssymb}
\usepackage{amsmath}
\usepackage{graphicx}
\usepackage{lipsum}
\usepackage{stix}
\usepackage{siunitx}
\usepackage{mwe}
\usepackage{lineno}
\usepackage{siunitx}
\usepackage{url}
\usepackage{subfigure}
\usepackage[colorlinks=true]{hyperref}

\begin{document}
\begin{frontmatter}
\title{Neutron spectroscopy with a high-pressure nitrogen-filled spherical proportional counter}

\author[aff1]{I.~Giomataris}
\author[aff2]{S.~Green}
\author[aff2]{I.~Katsioulas}
\author[aff2]{P.~Knights}
\author[aff2]{I.~Manthos\corref{cor1}}
\ead{i.manthos@bham.ac.uk}
\author[aff2]{T.~Neep}
\author[aff2]{K.~Nikolopoulos}
\author[aff1]{T.~Papaevangelou}
\author[aff2]{B.~Phoenix}
\author[aff2]{J.~Sanders}
\author[aff2]{R.~Ward}

\address[aff1]{IRFU, CEA, Universite Paris-Saclay, F-91191 Gif-sur-Yvette, France}
\address[aff2]{School of Physics and Astronomy, University of Birmingham, B15 2TT, United Kingdom}

\journal{Journal of Nuclear Instruments and Methods in Physics Research A}

\cortext[cor1]{Corresponding author}

\begin{abstract}
The spherical proportional counter is a large volume gaseous detector which finds application in several fields, including direct Dark Matter searches. When the detector is filled with nitrogen it becomes an effective neutron spectrometer thanks to the $^{14}$N(n,$\mathrm{\alpha}$)$^{11}$B and $^{14}$N(n,p)$^{14}$C reactions. Nitrogen, however, is a challenging operating gas for proportional counters and requires a high electric field strength to gas pressure ratio. Benefiting from the latest advances in spherical proportional counter instrumentation and simulation techniques, we report first neutron measurements at operating pressures of up to 1.8\,bar.  This achievement enhances the prospects of the spherical proportional counter to act as a neutron spectrometer appropriate for challenging environments, including underground laboratories, and industrial and medical settings.
\end{abstract}


\end{frontmatter} 

\section{Introduction} \label{intro}
Neutron spectroscopy is an invaluable tool for a broad range of scientific, industrial, and medical applications~\cite{applications}; spanning from direct Dark Matter (DM) searches~\cite{PhysRevD.105.012002} and neutrino-less double-beta decay~\cite{Meregaglia:2017nhx} to non-destructive measurements of materials and medical imaging or cancer treatment.  For example, in underground laboratories, neutron induced backgrounds caused by cosmic ray muons and natural radioactivity may mimic the DM signal, reducing experimental sensitivity. A dedicated, precise, and in-situ measurement of neutron flux would be a valuable tool for characterising and mitigating neutron background. Amidst several attempts to develop an efficient neutron spectroscopy system, detection methods are complex, and measurements are scarce. Furthermore, despite significant effort invested in describing the neutron background with Monte Carlo simulations~\cite{montecarlo}, there is still significant uncertainty on the expected neutron background rates.

To date, neutron detection most widely relies on the $^3$He(n,p)$^3$H reaction, providing high efficiency for thermal neutrons and very low efficiency for $\gamma$-rays. However, $^3$He-based detectors only provide combined slow and fast neutron flux measurements, while energy measurements for fast neutrons are plagued by the wall effect, where the recoiling proton escapes the active volume. Solutions based on gas mixtures with heavy elements and operation at extremely high pressures are, typically, impractical. Moreover, the world-wide demand for $^3$He, which is largely driven by the popularity of such detectors, is in stark contrast to its supply resulting in a dramatic price increase. Thus, $^3$He-based detectors --especially larger ones-- require expensive retrieval systems, leading governments and industries in a search for alternatives.  Existing alternatives~\cite{cryst9090480,Simpson2011,KOUZES20101035,Kouzes:2015tsc}, have major disadvantages:
a)~Ga-based detectors, that exhibit the highest known cross section for thermal neutrons, present high-cost and long growing time, while require complex experimental setup for fast neutron spectroscopy;
b)~The gas in BF$_3$-based proportional counters is toxic and corrosive;
c)~Boron-lined tubes exhibit poor efficiency and high cost;
d)~Bulk scintillators have complicated response functions, insufficient $\gamma/$n discrimination, and limited radiation hardness, and;
e)~$^6$Li coated detectors filled with argon exhibit degraded energy resolution.

A novel approach to neutron spectroscopy for thermal and fast neutrons has been proposed in Ref.~\cite{neutron}, relying on the spherical proportional counter~\cite{spc},  a high-gain and large-volume gaseous detector, operated with nitrogen. It relies on the $^{14}$N(n,$\alpha$)$^{11}$B and $^{14}$N(n,p)$^{14}$C processes, which have cross-sections comparable to the $^3$He(n,p)$^3$H reaction for fast neutrons. Thus, this approach has the potential to become an inexpensive, robust, and reliable alternative to $^3$He-based detectors. Nevertheless, the challenge lies in achieving sufficient gas gain in nitrogen-filled proportional counters. As a result, early proof-of-principle demonstrations were limited to low-pressure measurements up to 0.5\;bar, which resulted in non-negligible wall effect.

In this work, we implement the latest developments in spherical proportional counter instrumentation, e.g. the multi-anode resistive sensor~\cite{achinos,achinos2}, to achieve fast and thermal neutron spectroscopy with a nitrogen-filled spherical proportional counter at pressures up to 1.8\,bar. Results from a dedicated simulation framework \cite{simulation} are also presented and compared to the experimental data.

\section{Spherical Proportional Counter}

\begin{figure}[h]
\centering
\subfigure[\label{fig:fig1a}]{\includegraphics[width=0.40\linewidth]{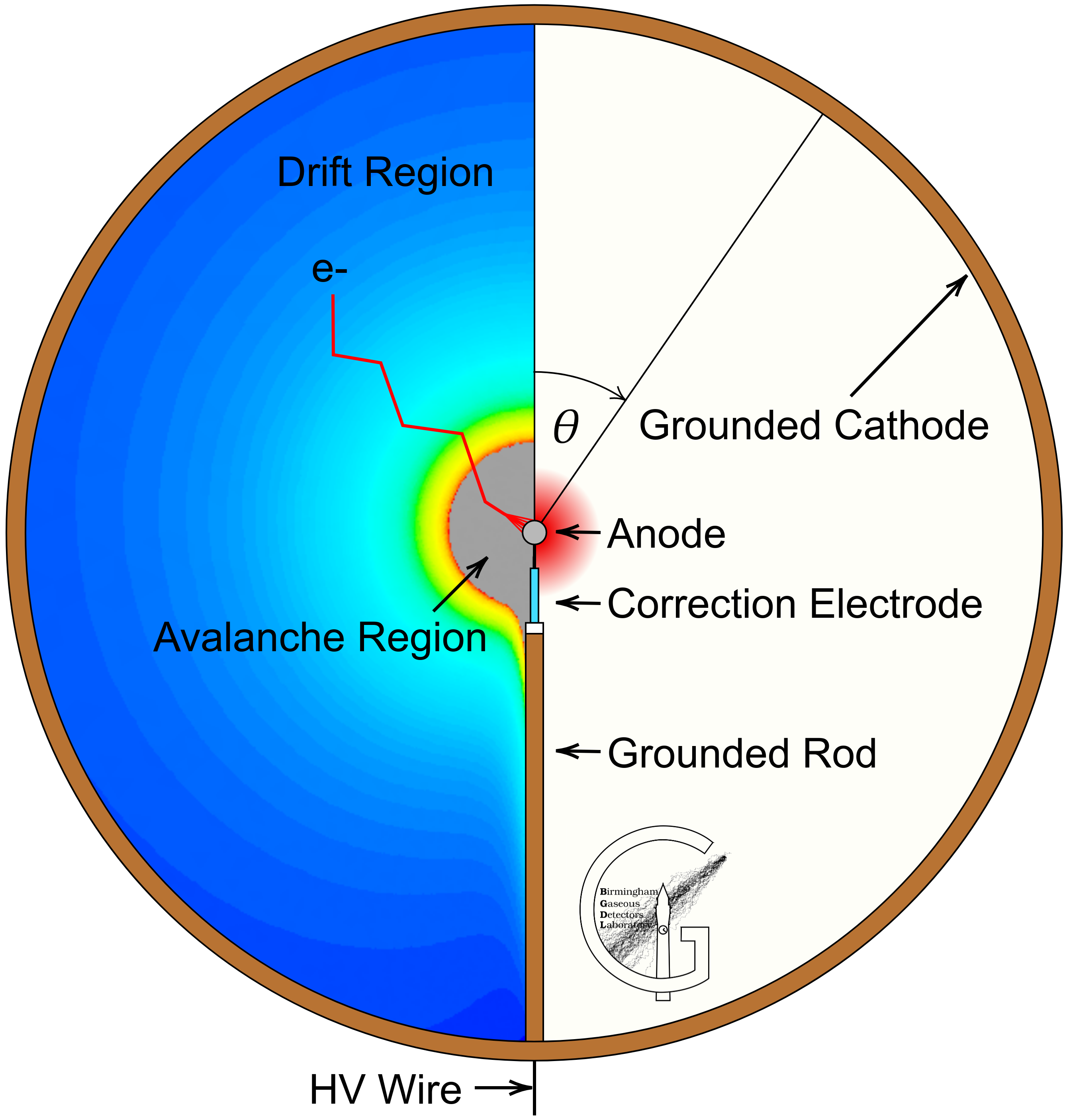}}
\subfigure[\label{fig:fig1b}]{\includegraphics[width=0.43\linewidth]{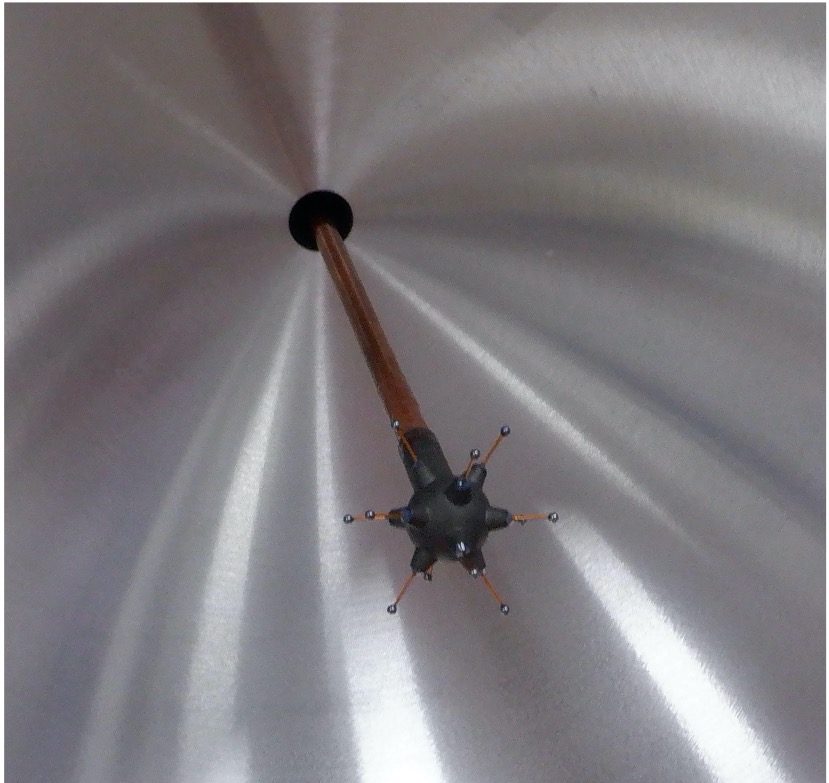}}
\caption{\subref{fig:fig1a} Spherical proportional counter schematic. \subref{fig:fig1b} 11-anode multi-anode ACHINOS sensor.\label{fig1}}
\end{figure}
The spherical proportional counter is presented in Figure~\ref{fig:fig1a}. It is a gaseous detector where a grounded spherical vessel serves as the cathode and a small spherical anode is placed at its centre. The operating high voltage is supplied to the anode through a wire, which is shielded by a grounded metallic rod that also supports the anode. In the ideal case, the electric field in the spherical proportional counter is radial and its magnitude can be approximated as $E(r)\approx \frac{V_{0}}{r^{2}} r_{a}$,
where $V_{0}$ is the anode voltage, $r_a$ the anode radius, and $r$ the distance from the detector centre. As a result the detector volume is separated into a drift and amplification region. 

The detector operation is governed by $E/P$, the ratio of the electric field strength to the gas pressure $P$. For spherical proportional counters operating with nitrogen, a high $E/P$ is required to achieve sufficient gas gain.
This is achieved with the multi-anode sensor ACHINOS~\cite{achinos, achinos2}, shown in Figure \ref{fig:fig1b}. ACHINOS comprises a spherical central resistive electrode with anodes arranged in a regular pattern. The anodes are placed at a constant distance from the central electrode, lying on the surface of a virtual sphere. With ACHINOS, the drift and amplification electric field strengths are decoupled: the drift region electric field is determined by the collective field of all the anodes, while the amplification region electric field is determined by the anode radius. In this way, larger $E/P$ values can be obtained in the drift region without unacceptably large $E/P$ values in the amplification region.

\section{Neutron spectroscopy with nitrogen-filled spherical proportional counter} 
Neutron detection with a nitrogen-filled spherical proportional counter relies on two neutron absorption reactions:
\begin{eqnarray*}
^{14}\textrm{N} + \textrm{n} \rightarrow {}^{14}\textrm{C} + \textrm{p} + 625\,{\textrm{keV}}\\
^{14}\textrm{N} + \textrm{n} \rightarrow {}^{11}\textrm{B} + \alpha - 159\,{\textrm{keV}}\,.
\end{eqnarray*}
The first process has significant cross-section for thermal neutrons and fast neutrons with energies up to 2\,MeV, while the latter has large cross-section for fast neutrons with energies above 2\,MeV, which is comparable to that of the $^3$He detectors. This is shown in Figure \ref{fig:fig2a}, where the cross-section of the two reactions is given as a function of the neutron energy, and it is compared with $^3$He(n,p)$^3$H, the process exploited by $^{3}$He-based detectors. Thus, nitrogen-filled spherical proportional counters are highly efficient for fast neutron detection without the need for moderation. The smaller cross-section of the $^{14}$N(n,p)$^{14}$C process for thermal neutrons with respect to the $^3$He-based detectors can be compensated by the possibility to construct larger volume detectors operated at high-pressures. Any potential wall effect, which is an issue for $^3$He-based detectors, is significantly suppressed due to the higher atomic number of nitrogen. At the same time, the detector has low sensitivity to $\gamma$-rays, the gas is inert and harmless and can be readily deployed at sites with strict safety requirements, while it is abundantly available at an accessible cost.

During the proof-of-principle demonstration of the method~\cite{neutron}, a 1.3\,m diameter spherical proportional counter was equipped with a 3\,mm diameter single-anode sensor and operated with nitrogen up to 500\,mbar. An anode voltage of 6.2\,kV was required to achieve sufficient gas gain.  Further increase of the anode voltage was prohibited by the sensor technology at the time, and thus it was not possible to explore higher-pressure operation. In this work, the use of the multi-anode sensor ACHINOS with 11 anodes of 1\,mm diameter enables operation with nitrogen at pressures up to 1.8\,bar, while maintaining the required anode voltage below 6\,kV.

\section{Experimental setup and detector characterisation}

\begin{figure}[h]
\centering
\subfigure[\label{fig:fig2a}]{\includegraphics[width=0.44\linewidth,height=11pc]{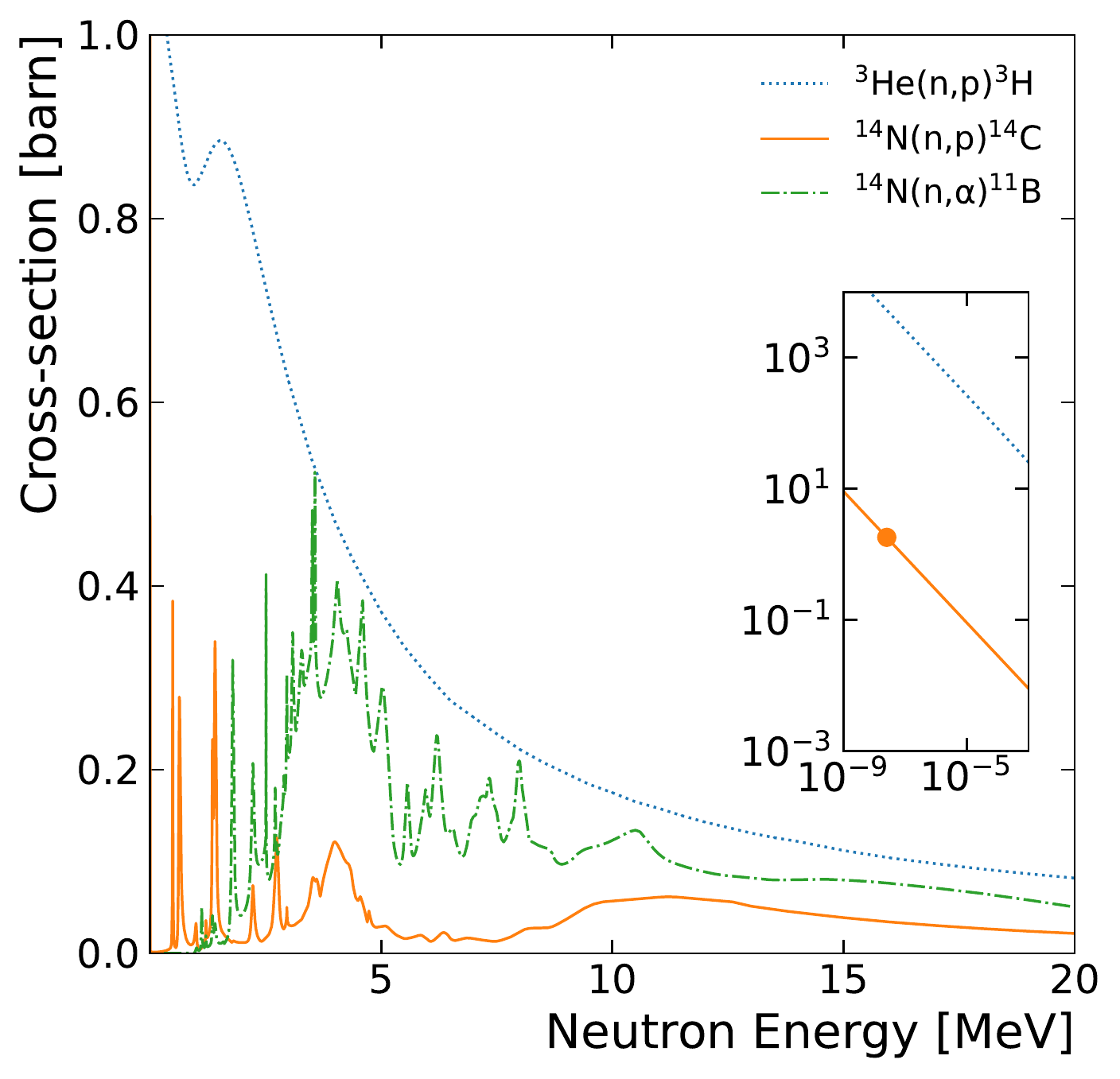}}
\subfigure[\label{fig:fig2b}]{\includegraphics[width=0.54\linewidth,height=11pc]{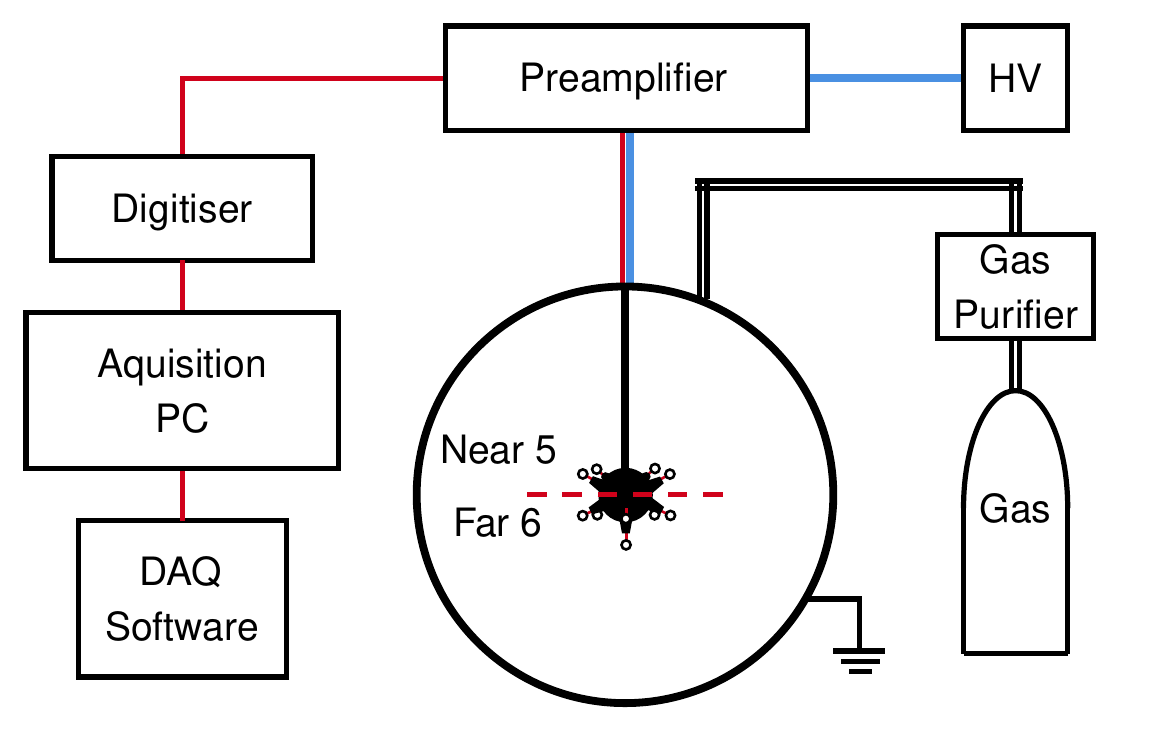}}
\caption{\subref{fig:fig2a} Neutron interaction cross-section as a function of the neutron kinetic energy. The inset shows the cross-sections at low energies with the thermal neutron cross-section of interest shown by the orange point (solid line). \subref{fig:fig2b} Diagram of the experimental setup.\label{fig2}}
\end{figure}

For the measurements presented below, a spherical proportional counter installed, calibrated, and operated at the University of Birmingham (UoB) is used. The experimental configuration is shown in Figure~\ref{fig:fig2b}. The detector comprises a 30\,cm diameter stainless steel vessel, equipped with an 11-anode ACHINOS, with 1\,mm diameter stainless steel spherical anodes, placed at 1.7\,cm from the central electrode. A two-channel readout is implemented, where the ``Near'' and ``Far'' channels correspond to the 5 anodes closest and the 6 anodes farthest from the grounded rod, respectively, as shown in Figure~\ref{fig:fig2b}~\cite{achinos}. The signals are fed to two ORTEC 142AH preamplifiers, through which the anode voltage (HV) is also provided. The signal digitisation is performed by a ``CALI-box'' 16-bit analogue-to-digital converter (ADC) constructed by CEA-Saclay~\cite{calibox} with a dynamic range of $\pm$1.25\,V and a maximum sampling frequency of 5\,MHz. The digitised signal is subsequently sent to the acquisition PC operating the SAMBA data acquisition (DAQ) software~\cite{samba}.

\begin{figure}[h]
\centering
\subfigure[\label{fig:fig3a}]{\includegraphics[width=0.48\linewidth, trim={0cm -7cm -7cm 18cm},clip]{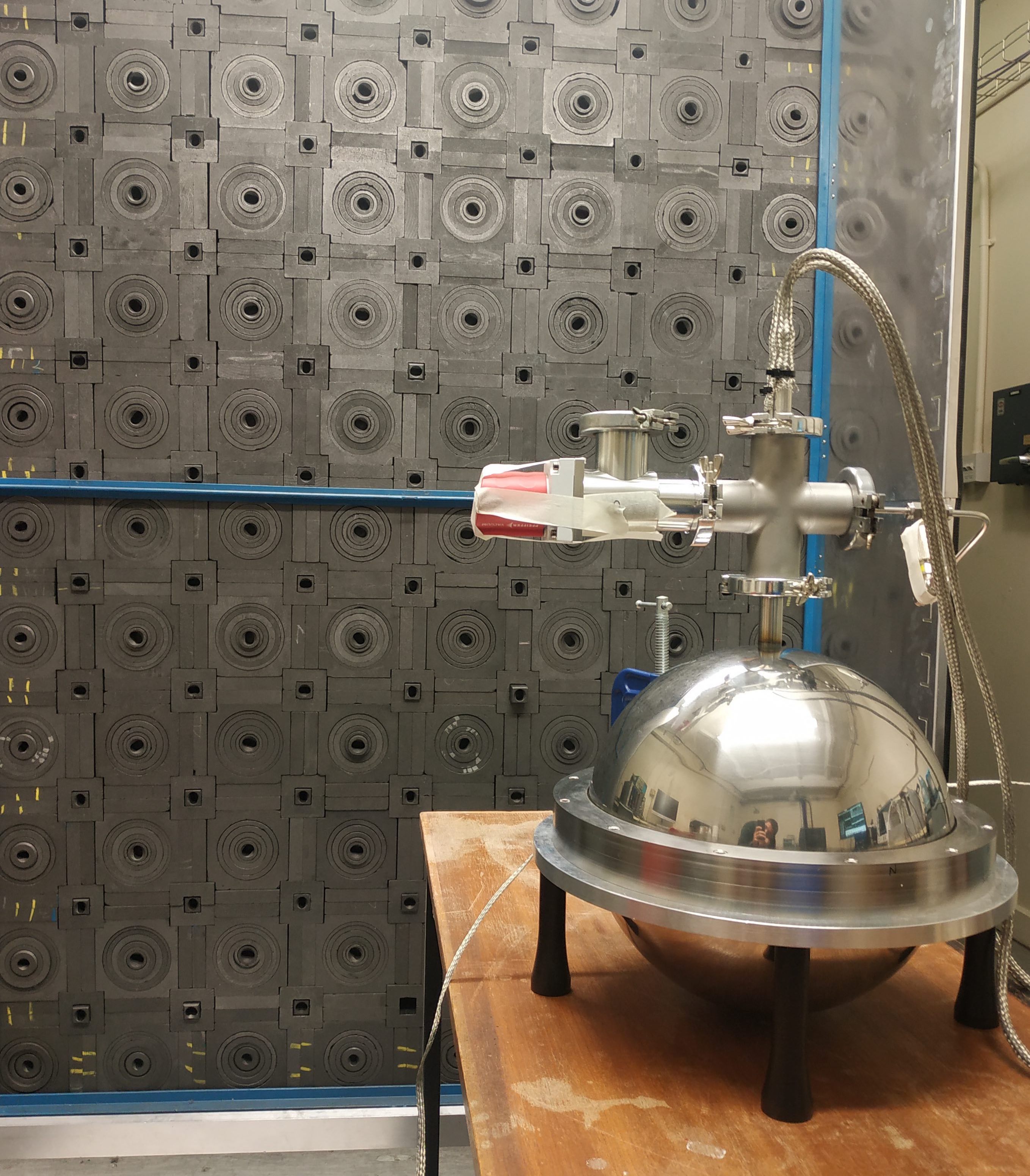}}
\subfigure[\label{fig:fig3b}]{\includegraphics[width=0.48\linewidth]{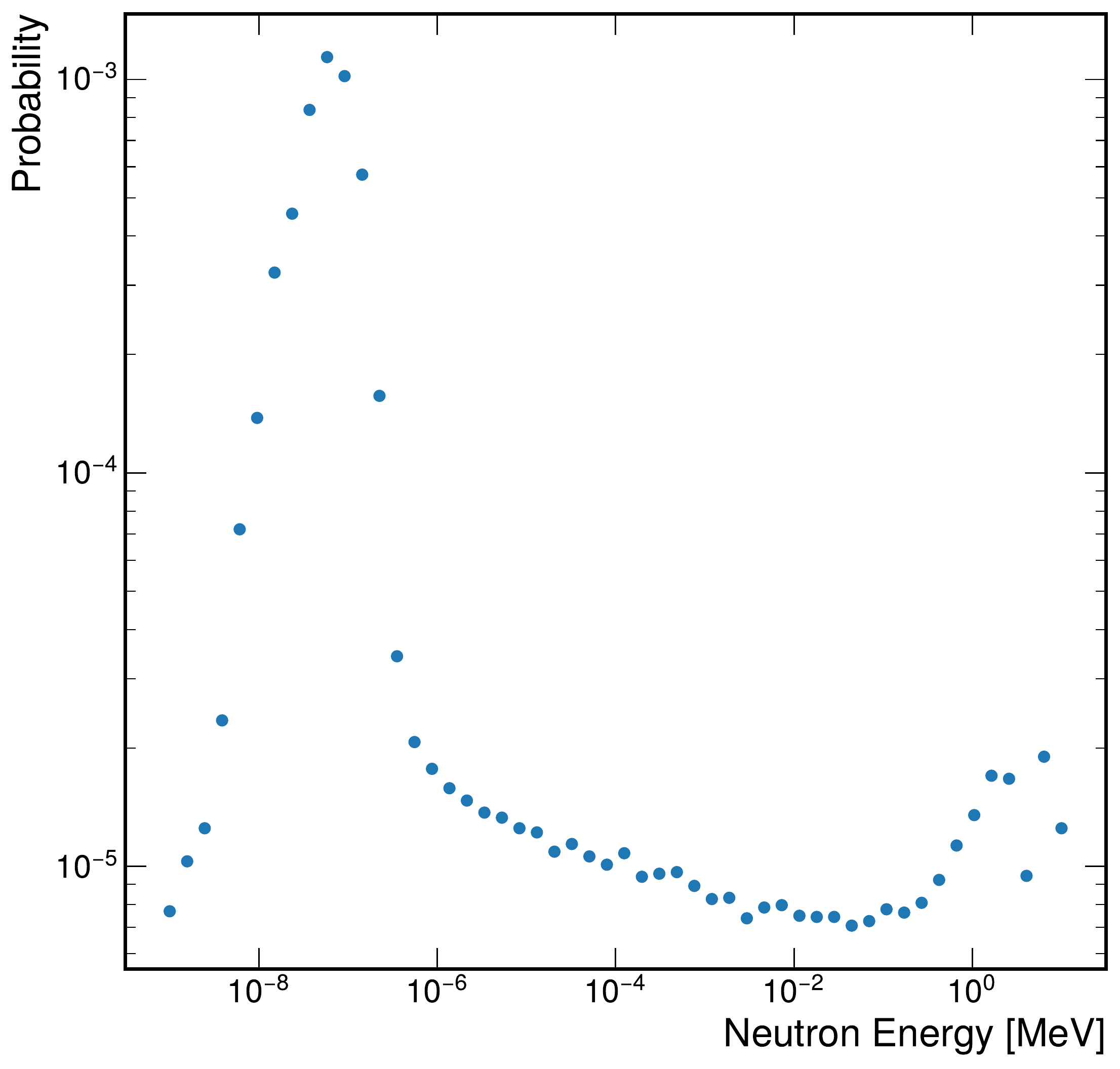}}
\caption{\subref{fig:fig3a} The spherical proportional counter at the graphite stack room at the University of Birmingham. \subref{fig:fig3b} Probability for a neutron from the $^{241}$Am-$^{9}$Be source to reach the detector volume as a function of its energy 
at the detector, obtained with MCNP simulations~\cite{mcnp}. \label{fig3}}
\end{figure}

In the presented work the detector response is evaluated using thermal and fast neutrons from an $^{241}$Am-$^9$Be source~\cite{MARSH1995340}, with an effective activity of $2.6\,\textrm{MBq}$. Thermal neutrons were produced by placing the source inside the UoB graphite stack, shown in Figure \ref{fig:fig3a}. The set-up was simulated using the Monte Carlo N-Particle radiation transport code (MCNP 6.1)~\cite{mcnp}. In Figure~\ref{fig:fig3b}, the probability of a moderated neutron of a given energy to reach the gas volume of the spherical proportional counter is presented.

\begin{figure}[h]
\centering
\subfigure[\label{fig:fig4a}]{\includegraphics[width=0.48\linewidth]{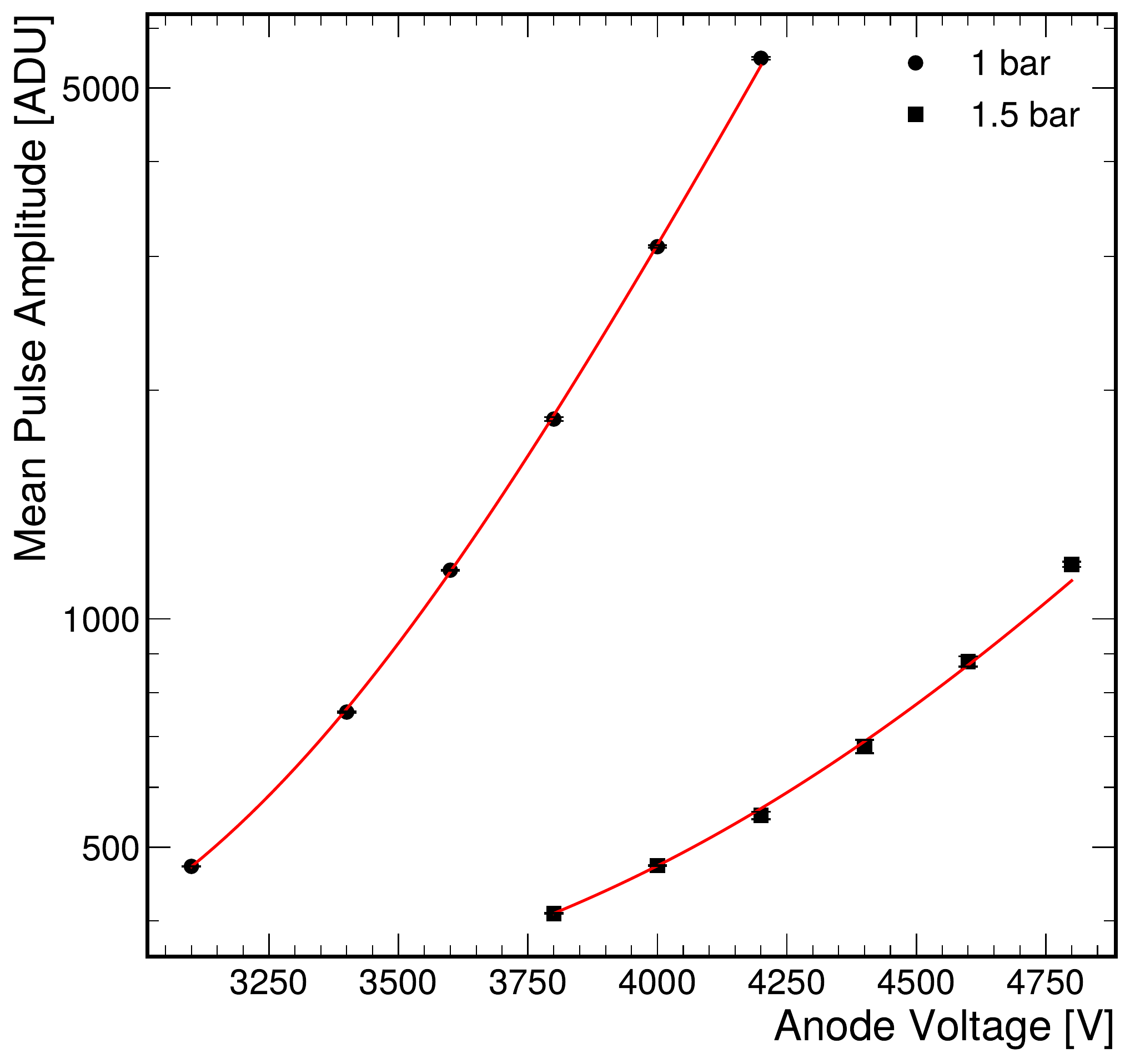}}
\subfigure[\label{fig:fig4b}]{\includegraphics[width=0.48\linewidth]{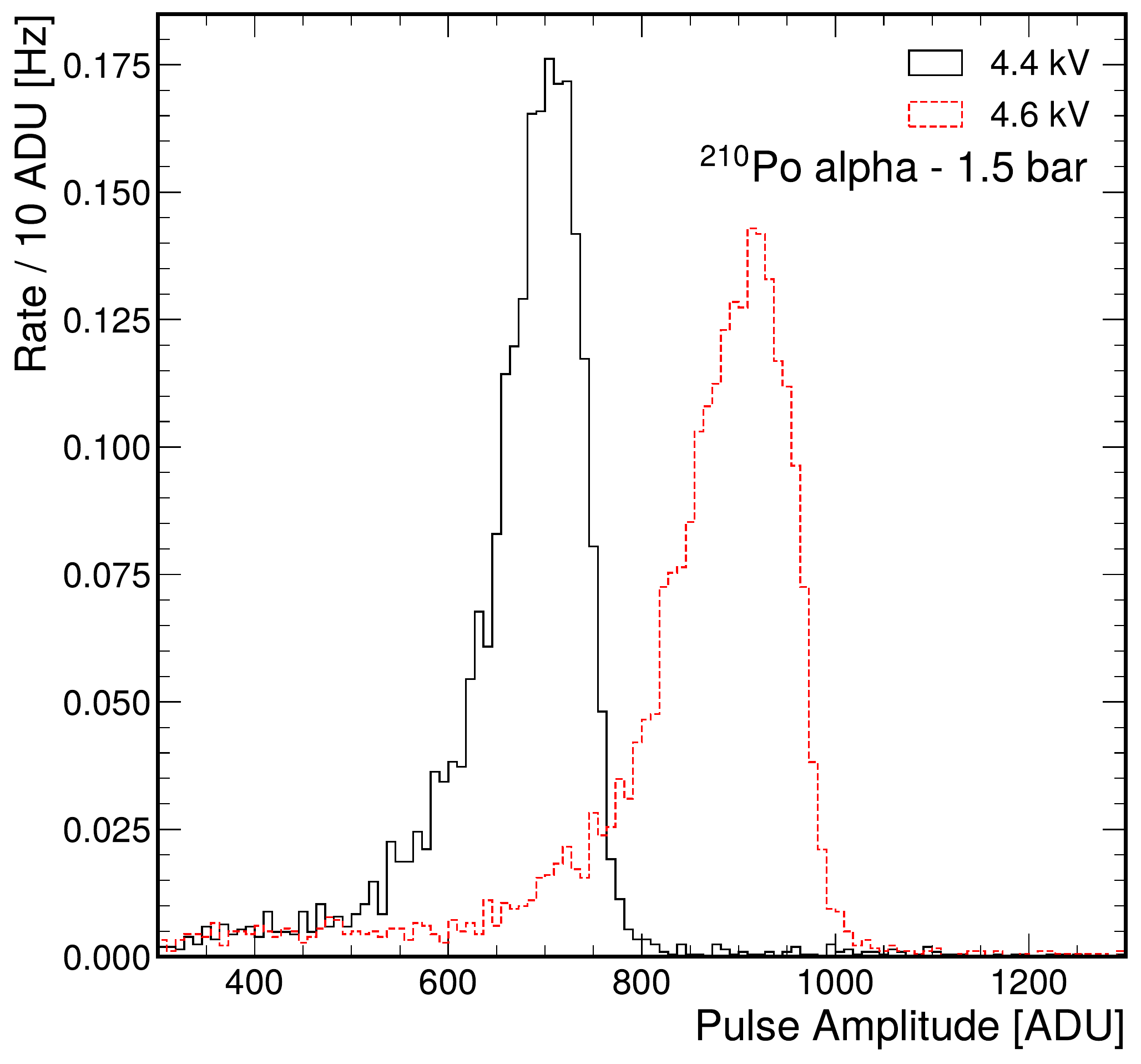}}
\caption{\subref{fig:fig4a} Gas amplitude curve with 5.30\,MeV $\alpha$-particles from $^{210}$Po decays with 1 and 1.5\,bar N$_2$. Estimated uncertainties are smaller than the marker size. 
\subref{fig:fig4b} Pulse amplitude distribution on the response of the spherical proportional counter to 5.30\,MeV $\alpha$-particles from $^{210}$Po decay at different anode voltages.
\label{fig:fig4}}
\end{figure}

The detector response was characterised using two approaches: a) A $^{210}$Po source and b) A gaseous $^{222}$Rn source. In the first case, a $^{210}$Po source emitting 5.30\,MeV $\alpha$-particles was placed on the inner surface of the cathode. For measurements at 1 and 1.5\,bar N$_2$ the source was only used for dedicated calibration runs, while for measurements at  1.8\,bar the $^{210}$Po source was continuously  in the detector providing a direct calibration during neutron runs. Two sets of measurements with the spherical proportional counter filled with 1 and 1.5\,bar N$_2$ at different anode voltages, provide the opportunity to extract the amplitude curve of the detector, fitted with an exponential and shown in Figure~\ref{fig:fig4a}. An example of two such measurements is shown in Figure~\ref{fig:fig4b} for 1.5\,bar pressure. Such curves inform the choice of the operating anode voltage, so that the energies from 625\,keV (corresponds to the recoil energy of thermal neutrons) to 20\,MeV is within the dynamic range of the electronics chain. 

In the second case  the detector was calibrated using the $^{222}$Rn emanated by a  commercial filter (Entegris MC700 902-F) used to remove H$_2$O and O$_2$ impurities from the gas. This is relevant for neutron measurements at 1 and 1.5\,bar  N$_2$. The $^{222}$Rn decay chain results in three distinct peaks at  5.59\,MeV, 6.11\,MeV, and 7.83\,MeV energies, produced by the $^{222}$Rn, $^{218}$Po and $^{214}$Po $\alpha$-decay, respectively.  The origin of these peaks was confirmed by a study of the time evolution of the event rate. The extracted half-life was within 3\% of the 3.82\,days expected for $^{222}$Rn. The gas purification for the measurements at 1.8\,bar was performed with a custom made filter~\cite{filter}, specifically designed to reduce the $^{222}$Rn emanation to low levels. Thus, in these measurements, the  $^{222}$Rn $\alpha$-particle rate was negligible compared to the neutron detection rate.

\section{Results}
The nitrogen-filled spherical proportional counter was operated at pressures up to 1.8\,bar with various anode voltages.
\begin{figure}[h]
\centering
\includegraphics[width=0.5\linewidth]{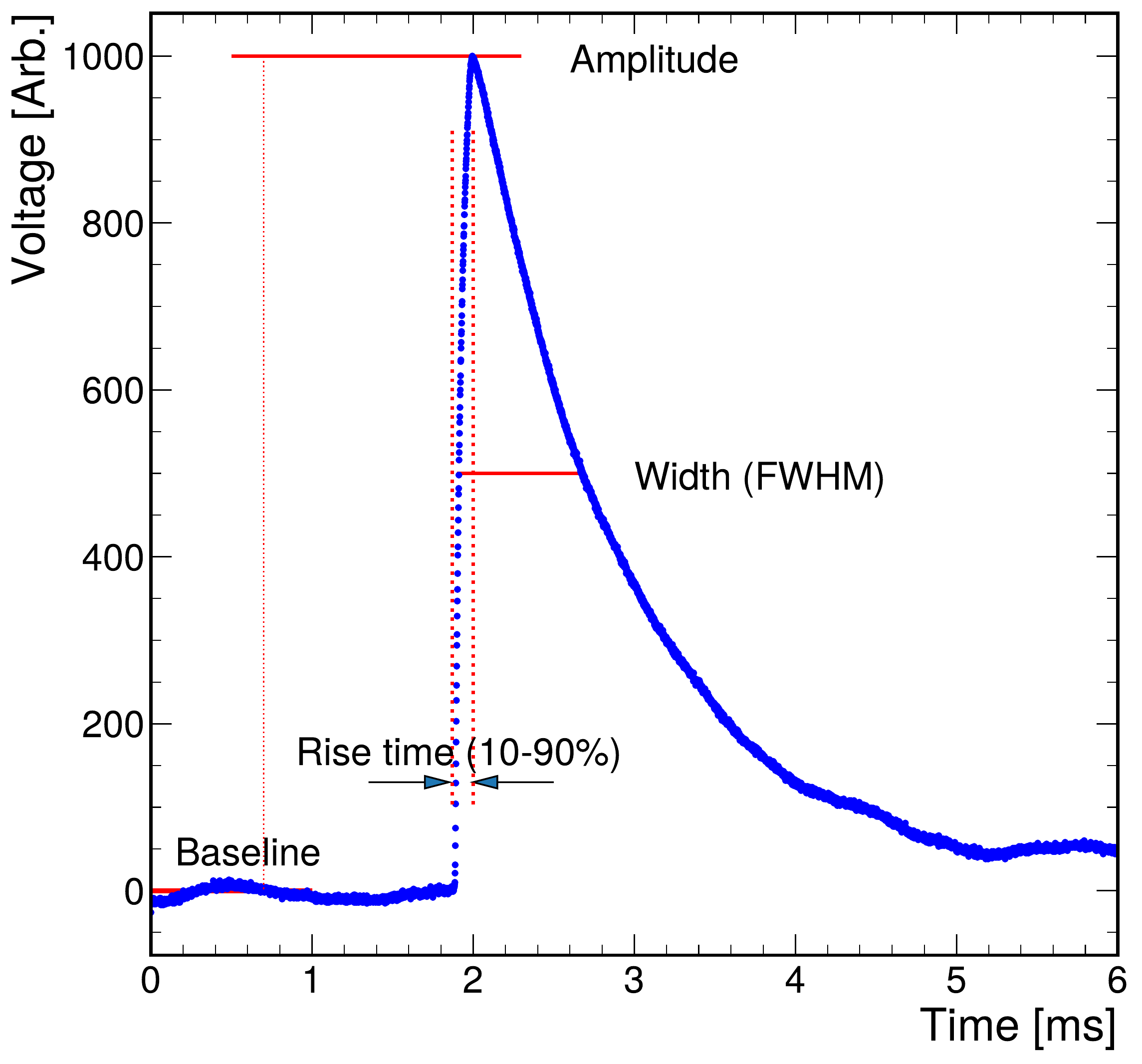}
\caption{A typical experimental pulse of the spherical proportional counter. The amplitude, width and rise time parameters used in the event selection, are indicated.\label{fig:fig5}}
\end{figure}
\begin{figure}[h]
\centering
\subfigure[\label{fig:fig6a}]{\includegraphics[width=0.32\linewidth]{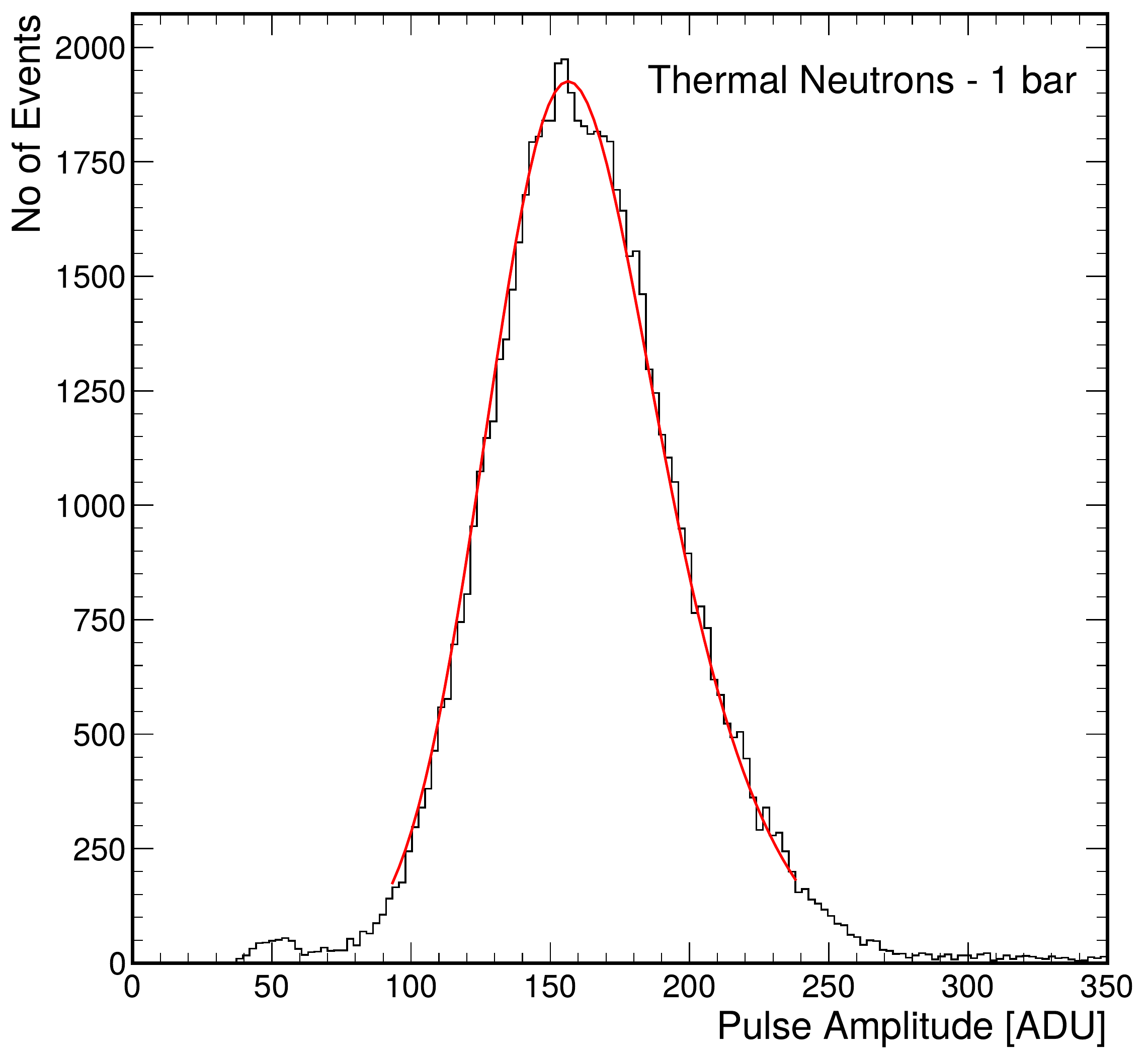}}
\subfigure[\label{fig:fig6b}]{\includegraphics[width=0.32\linewidth]{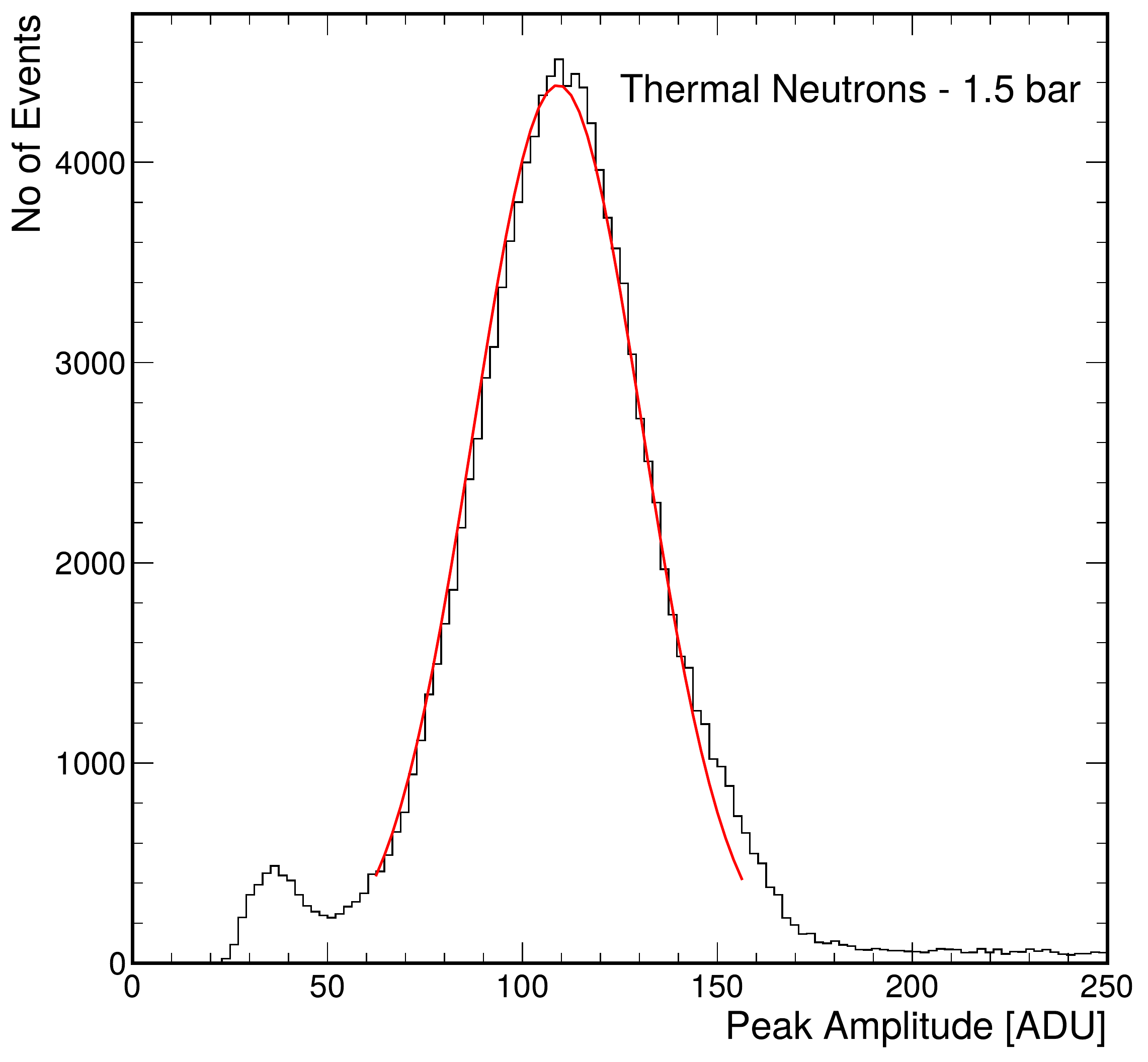}}
\subfigure[\label{fig:fig6c}]{\includegraphics[width=0.32\linewidth]{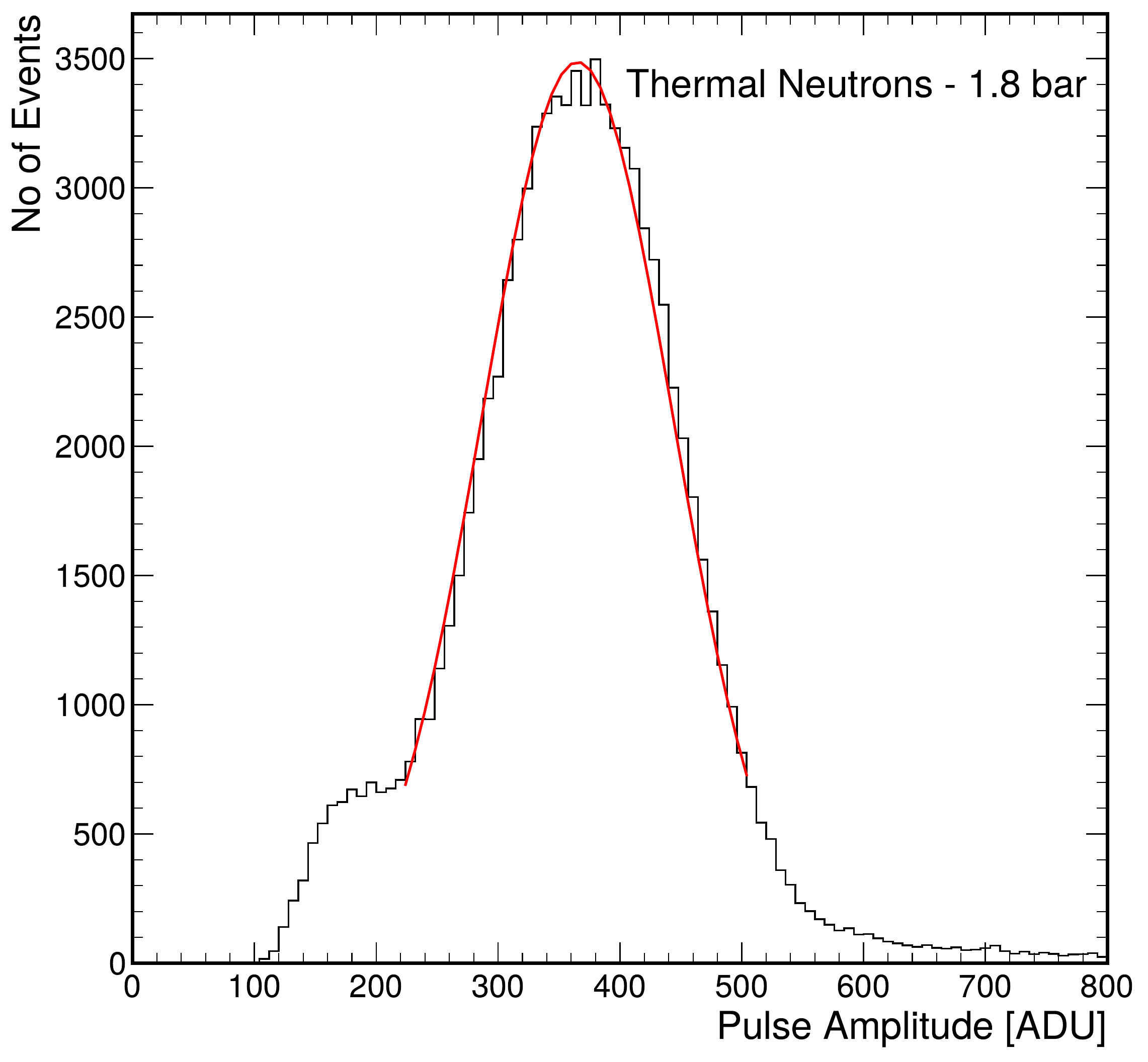}}
\caption{Thermal neutron detection at different nitrogen pressures: \subref{fig:fig6a} 1\,bar of nitrogen and anode voltage of 3.6\,kV, \subref{fig:fig6b} 1.5\,bar of nitrogen and bias voltage of 4.5\,kV,  and \subref{fig:fig6c} 1.8\,bar of nitrogen and anode voltage of 5.95\,kV. The peak at 162.1$\pm$6.0, 109.2$\pm$0.2 and 364.1$\pm$0.4\,ADU correspondingly defined by Gaussian fit of the amplitude distribution. The small peak at the left of each histogram is attributed to residual noise. \label{fig:fig6}}
\end{figure}
To distinguish neutrons detected in the gas volume from events originating from other sources, e.g. electronic noise or neutrons interacting at the cathode, standard pulse shape parameters were taken into account. Specifically,  the rise time, defined as the time required from the signal to increase from 10\% to 90\% of the pulse amplitude, and the pulse full-width at half-maximum (FWHM) were used. These are described in Figure~\ref{fig:fig5}. The signal pulse is expected to have a rise time of a few tens of $\si{\micro \textrm{s}}$ and provides information about the energy deposited in the detector volume through pulse size parameters (peak amplitude, integral of the pulse).

\begin{figure}[h]
\centering
\includegraphics[width=13.5pc]{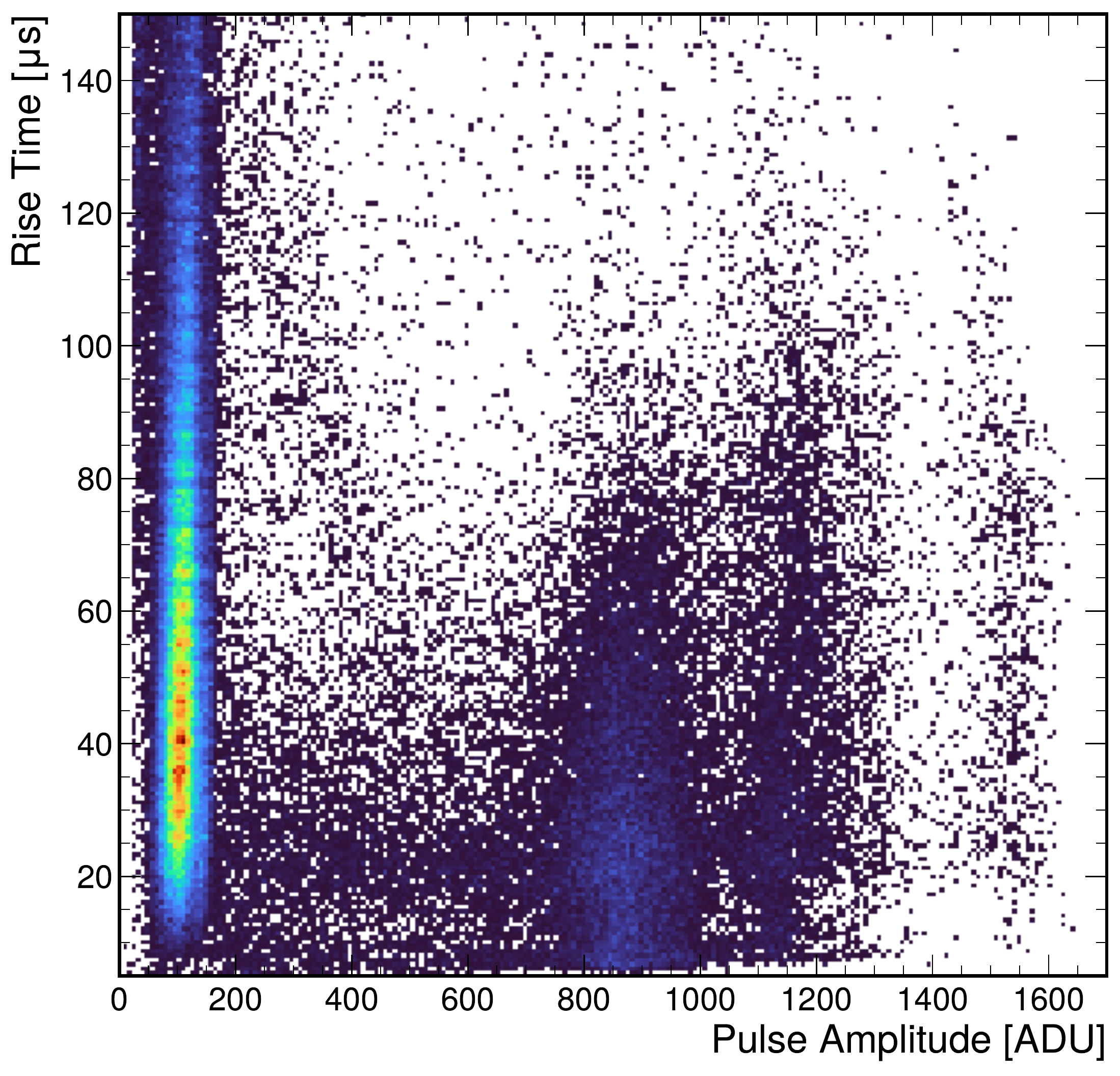}
\caption{Pulse rise time versus amplitude, showing thermal neutrons events with 1.5\,bar N$_{2}$ and 4.5\,kV at the anode at approximately 110\,ADU. Additional to the thermal neutrons, the $^{222}$Rn decay chain is shown, ranging from 800 to 1600\,ADU.\label{fig:fig7}}
\end{figure}
As a demonstration of thermal neutron detection, results with the detector filled with N$_2$ at 1, 1.5 and 1.8\,bar and operating at 3.6, 4.5 and 5.95\,kV anode voltage, are presented in Figure~\ref{fig:fig6}. 
The rise time versus the amplitude for the case of 1.5\,bar pressure is presented in Figure \ref{fig:fig7}. The concentration of events in the leftmost part of the plot is attributed to thermal neutrons, while the $^{222}$Rn decay chain is apparent at higher amplitudes. The relative positions of the peaks from the decay chain of $^{222}$Rn confirms the observation of the thermal neutron peak.
\begin{figure}[h]
\centering
\includegraphics[width=14pc]{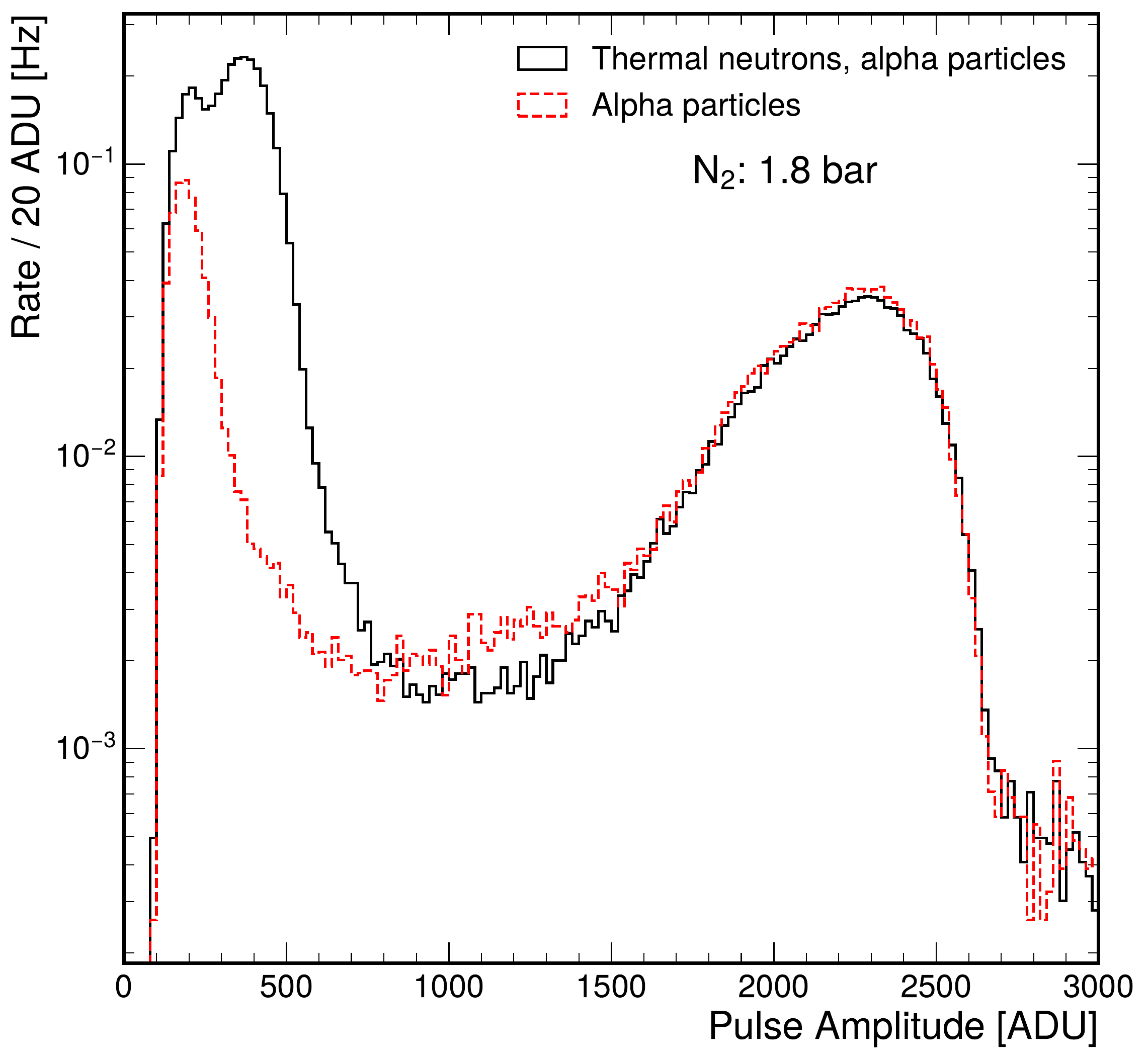}
\caption{Amplitude distribution with the spherical proportional counter operating with 1.8\,bar N$_{2}$ and 5.95\,kV  anode voltage. The peak with mean value 2176$\pm$41\,ADU corresponds to the $\alpha$-particles  from the $^{210}$Po sample providing the energy reference. The run with thermal neutrons  is presented with the solid line, while the run with the calibration source alone is shown with the dashed.\label{fig:fig8}}
\end{figure}
\begin{figure}[h]
\centering
\includegraphics[width=0.48\linewidth]{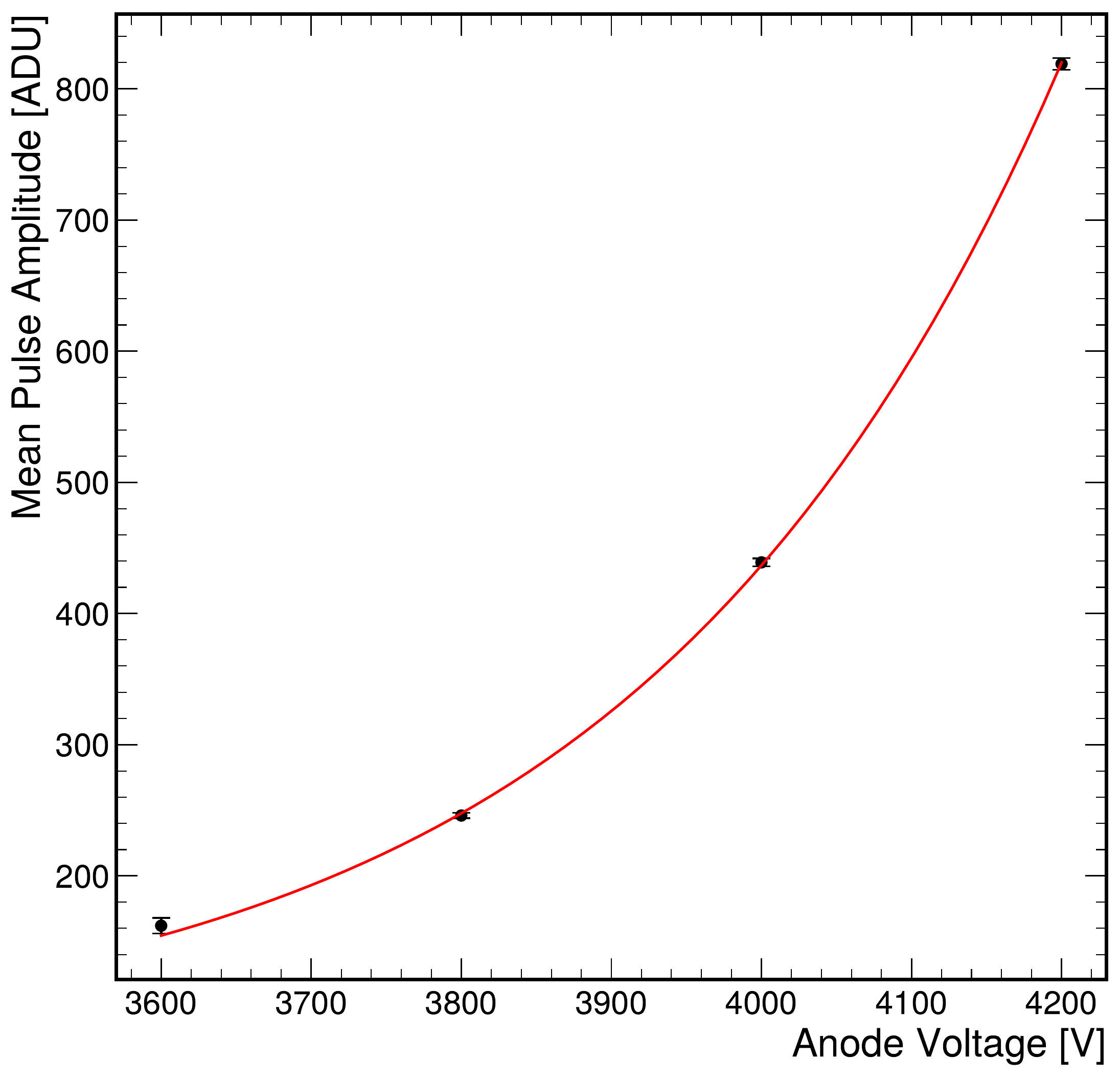}
\caption{Mean amplitude of the thermal neutron peak at 1\,bar as a function of  the anode voltage.  The error bars correspond to the standard deviation of the Gauss fit. An exponential fit is shown in red.\label{fig:fig9}}
\end{figure}
\begin{figure}[h]
\centering
\subfigure[\label{fig:fig10a}]{\includegraphics[width=0.475\linewidth]{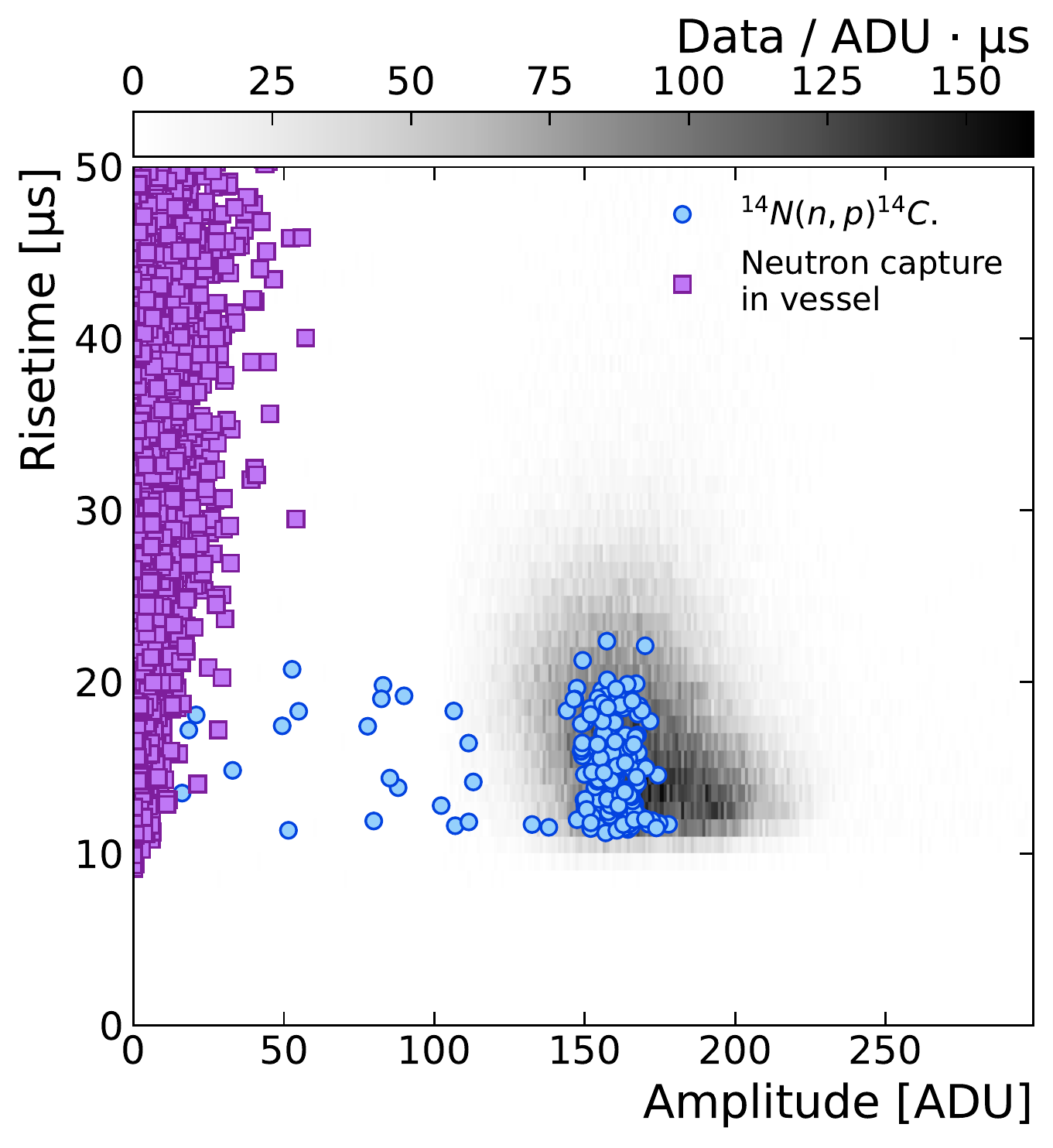}}
\subfigure[\label{fig:fig10b}]{\includegraphics[width=0.48\linewidth]{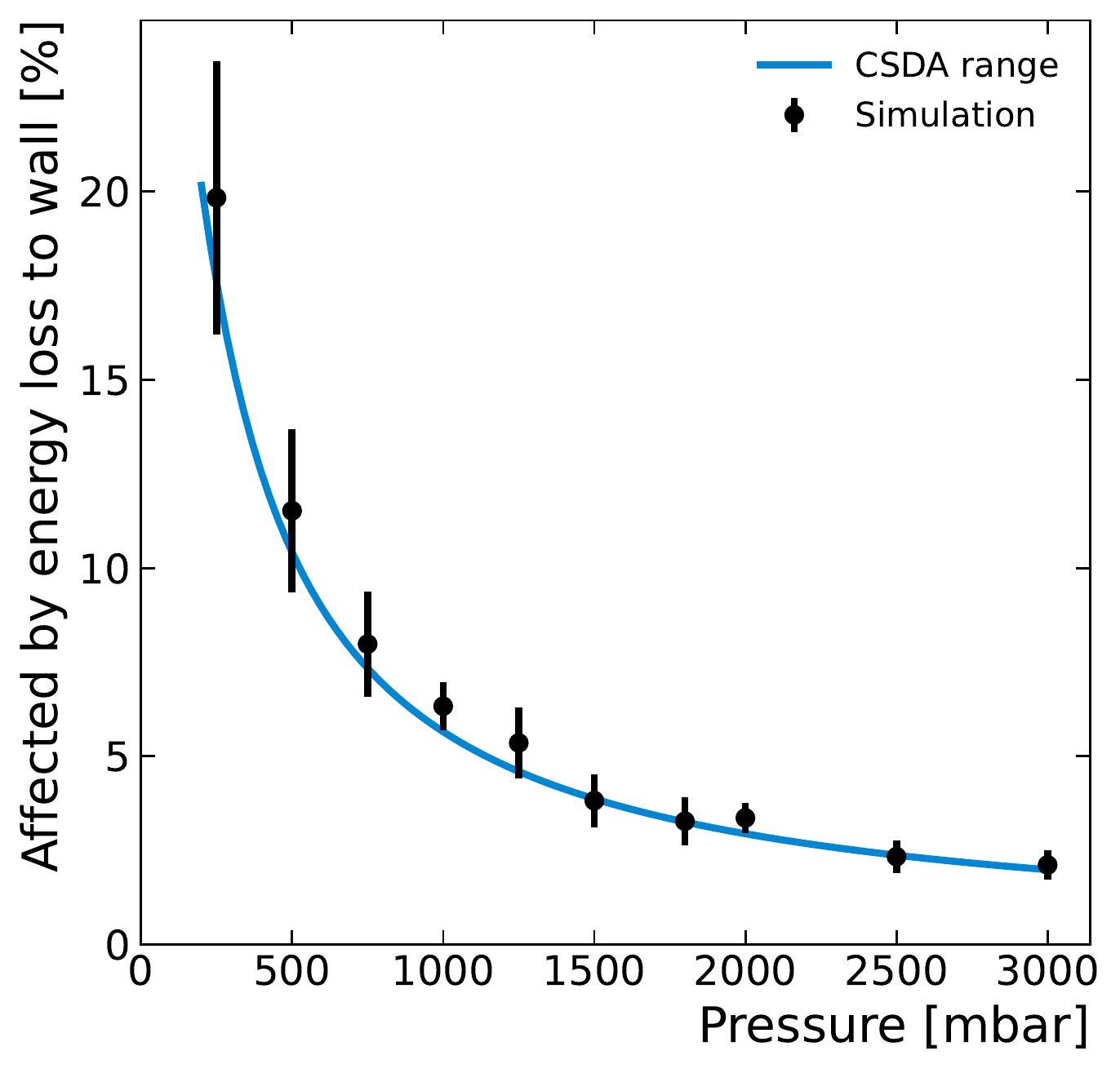}}
\caption{\subref{fig:fig10a} Pulse amplitude versus rise time from simulation (coloured circles/squares) compared with data (histogram). There is good agreement between the prediction for the $^{14}$N(n,p)$^{14}$C process and the measured data. Events from neutron capture in the vessel have amplitudes lower than the trigger threshold used in data taking.  
\subref{fig:fig10b} The percentage of thermal neutron interactions in which energy is lost to the wall of the SPC. The black points are obtained using the simulation framework described in the text, with error bars representing statistical uncertainties. The blue line is the expected wall effect probability calculated using the CSDA range of a 625\,keV proton, taken from the \textsc{pstar} database~\cite{pstar}. \label{fig:fig10}}
\end{figure}
\begin{figure}[h]
\centering
\includegraphics[width=0.5\linewidth]{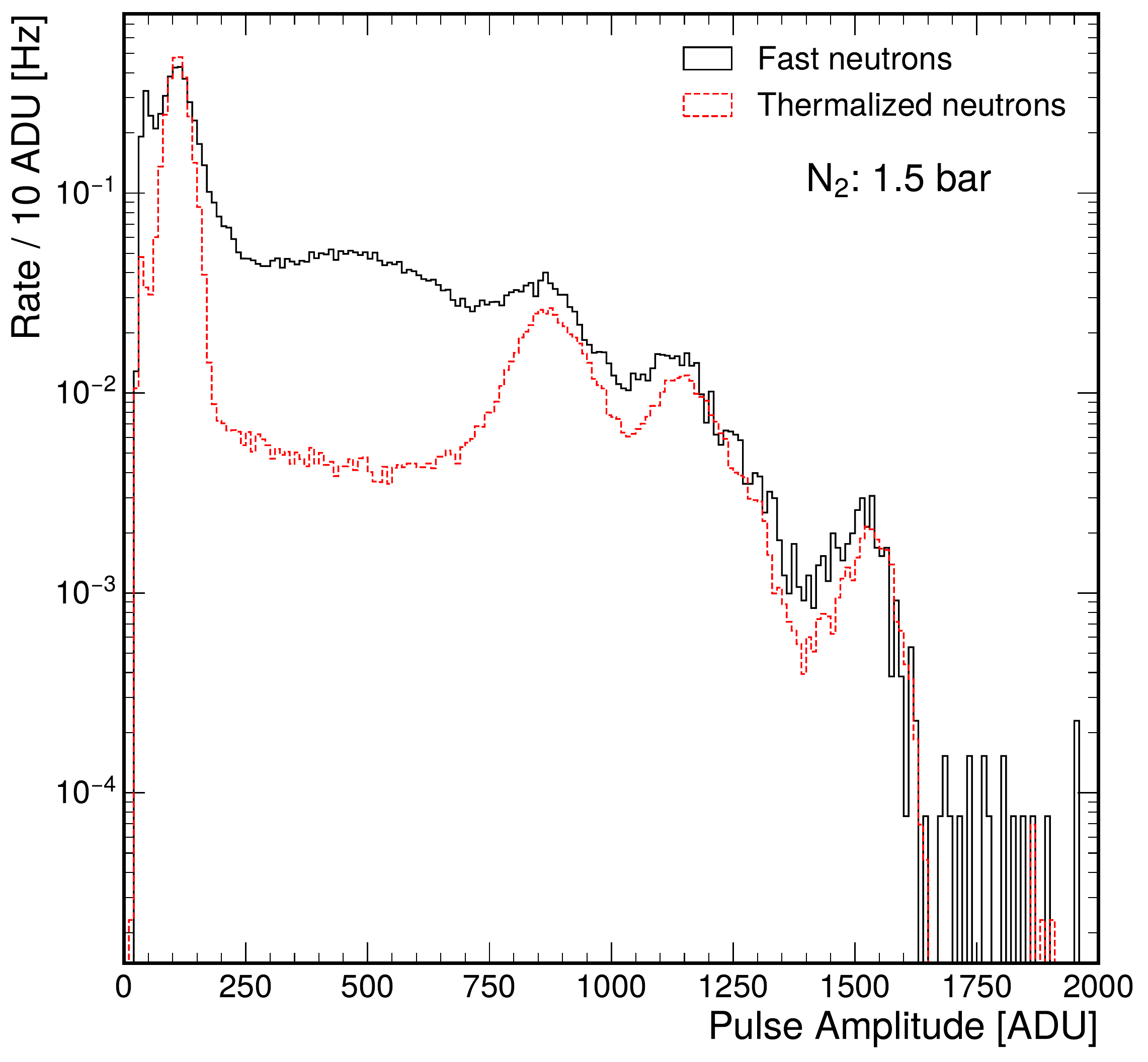}
\caption{Amplitude distribution of thermal neutrons in comparison to fast neutrons from the $^{241}$Am-$^{9}$Be source when operating with 1.5 bar N$_2$  and 4.5\,kV anode voltage. The three rightmost peaks correspond to $^{222}$Rn decay chain.\label{fig:fig11}}
\end{figure}
\begin{figure}[h]
\centering
\subfigure[\label{fig:fig12a}]{\includegraphics[width=14pc]{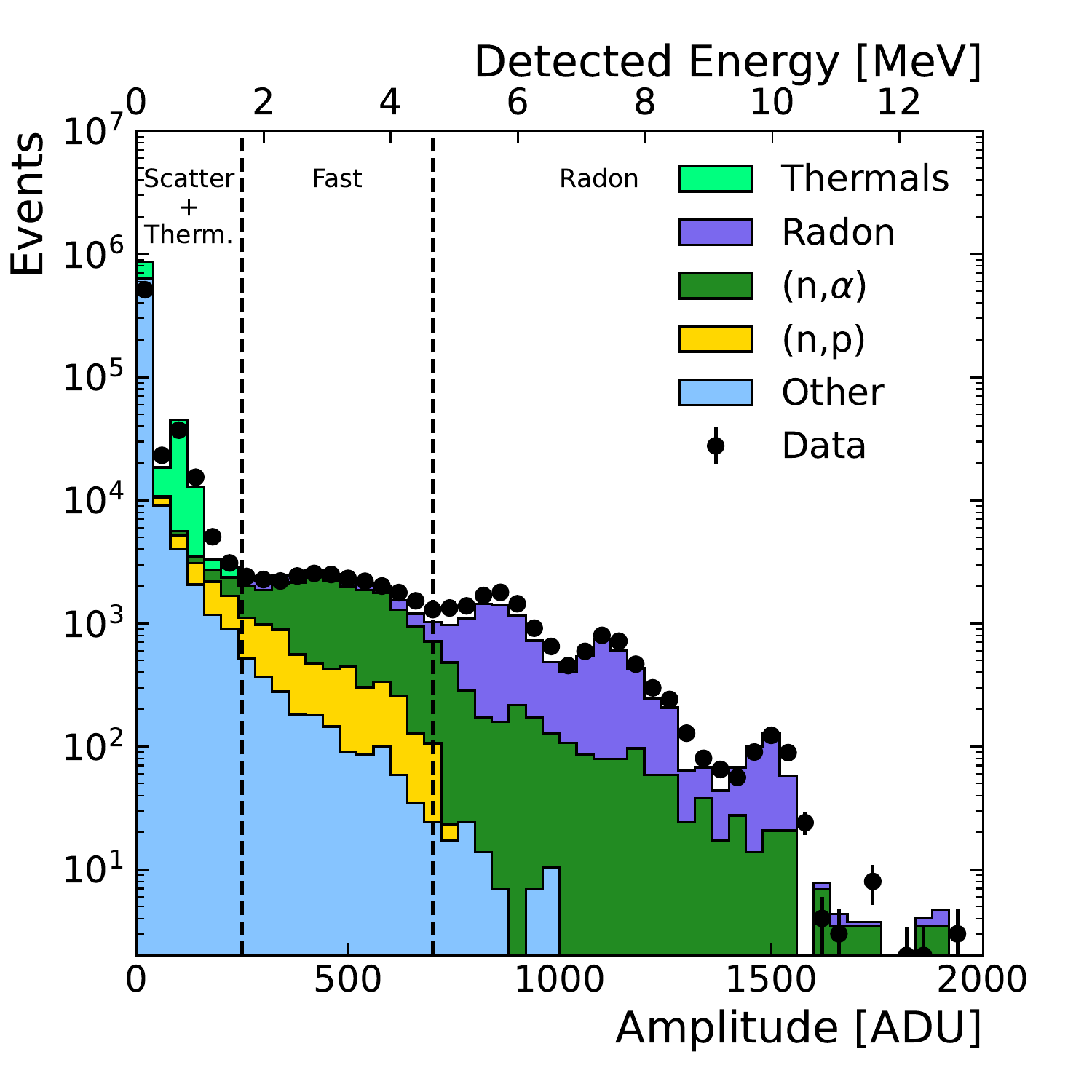}}
\subfigure[\label{fig:fig12b}]{\includegraphics[width=14pc]{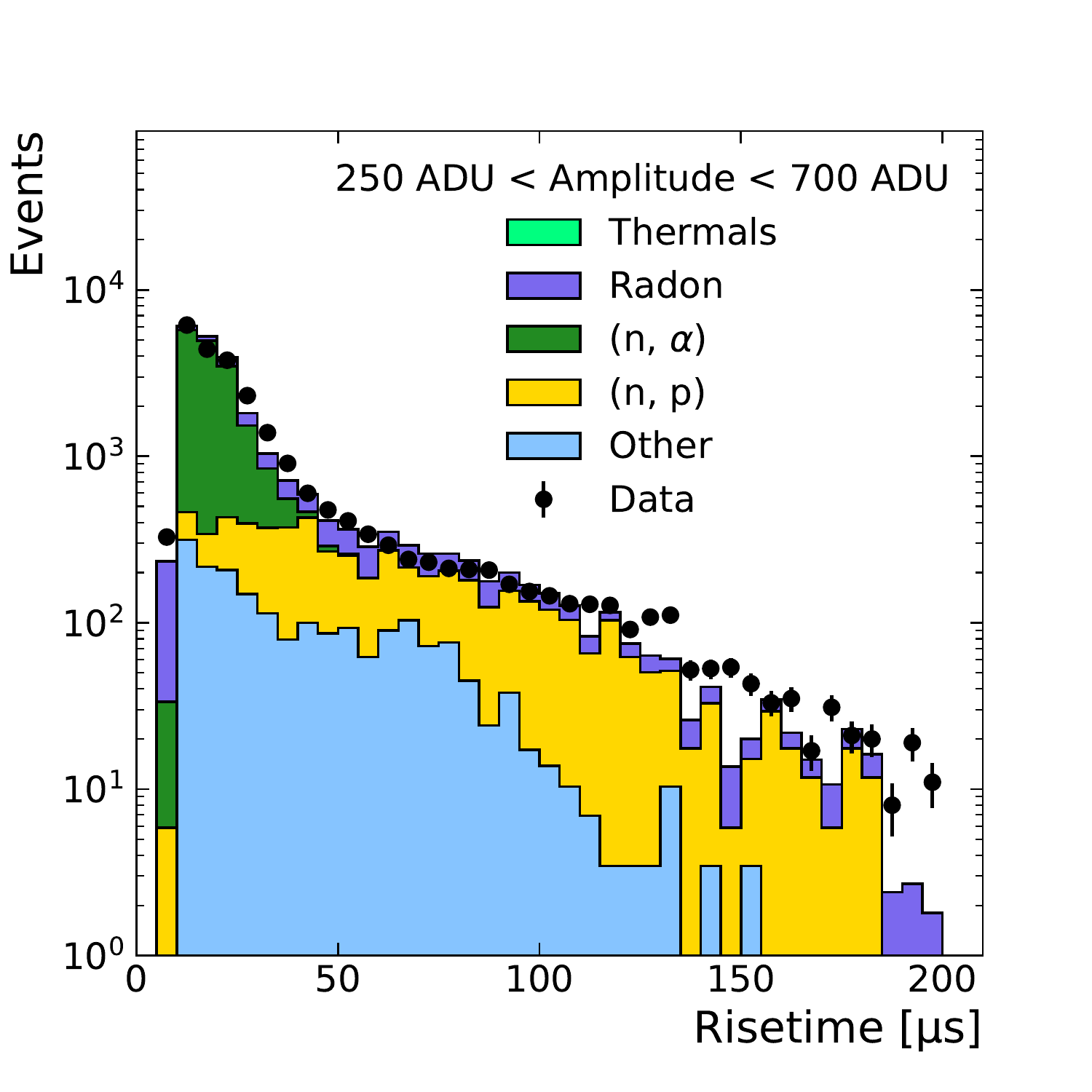}}
\caption{\subref{fig:fig12a} Fast neutrons pulse amplitude distribution from the $^{241}$Am-$^{9}$Be neutron source in comparison with the simulation results that indicate the origin of the detected events. \subref{fig:fig12b} Pulse rise time distribution for the events of Figure \ref{fig:fig12a} indicated as fast.  
\label{fig:fig12}}
\end{figure}

The response of the detector for neutron measurements at 1.8\,bar and 5.95\,kV anode voltage is presented in Figure \ref{fig:fig8}. The response of the spherical proportional counter to thermalised neutrons and to the $\alpha$-particles from the $^{210}$Po source is shown in black and the   background run with only the $^{210}$Po source is presented in red.
The $\alpha$-particles peak is fitted with the sum of two Gaussian, providing the energy reference and validating the thermal neutron peak shown in Figure~\ref{fig:fig6c}.

A compilation of the thermal neutron detection measurements at 1\,bar is shown in Figure \ref{fig:fig9}. The mean of the Gaussian function that fits the peak amplitude distribution is presented as a function of the anode voltage. The uncertainty represents the standard deviation of the Gaussian fit, while the response of the spherical proportional counter to thermal neutrons follows an exponential dependence with the applied anode voltage, in agreement with the amplitude curve found during detector characterisation.

The spherical proportional counter response was studied with the use of the UoB simulation framework \cite{simulation}, which combines Geant4 \cite{geant4} for the simulation of the passage of particles through matter, Garfield++ \cite{garfield} for the interactions of particles with the gas and the signal formation. The electric field is described with the use of ANSYS \cite{ansys} finite element method software. The response of the spherical proportional counter filled with 1\,bar of nitrogen to 3 million thermal neutrons isotropically incident on the detector was simulated. These simulated events were used to estimate the detection efficiency and to derive the expected pulse shape parameters for comparison with the data.
Combining  the 2.6\,MBq activity of the $^{241}$Am-$^{9}$Be neutron source, the 5$\times$10$^{-3}$ probability that a neutron emitted by the source reaches the detector volume --obtained by integrating the spectrum of Figure~\ref{fig:fig3b}-- and the detection rate  which was estimated to be 5.0$\pm$0.1\,Hz for all datasets at 1\,bar pressure, an efficiency of approximately 0.04\% for thermal neutrons was obtained. This is found to be in good agreement with the simulation that provides an efficiency $0.057\pm0.003 ~\mathrm{(stat.)}~\%$ for thermal neutron detection. At 1.8\,bar, the detection rate was measured to be 9.4\,Hz, corresponding to an efficiency of 0.08\%, while the simulation predicts an efficiency of $0.104\pm0.003~\mathrm{(stat.)}~\%$. An estimation of the fast neutron efficiency for the dataset at 1.5\,bar, accounting for the source activity, its position relative to the detector, and the detection rate for neutrons of higher energy than the thermal peak, provide 0.011\% efficiency, while simulation resulted to $0.018\pm0.001 ~\mathrm{(stat)}~\%$. This is in accordance with the expected proportional increase of the detector efficiency with pressure and the cross section difference between fast and thermal neutrons, as can be seen in Figure \ref{fig:fig2a}.
The pulse amplitude versus rise time obtained from the simulation is compared with the measured data in Figure~\ref{fig:fig10a} and provides insights on the different neutron interactions. Thermal neutron interactions via the $^{14}$N(n,p)$^{14}$C process within the detector volume are denoted by blue circles.
\begin{figure}[h]
\centering
\includegraphics[width=0.48\linewidth]{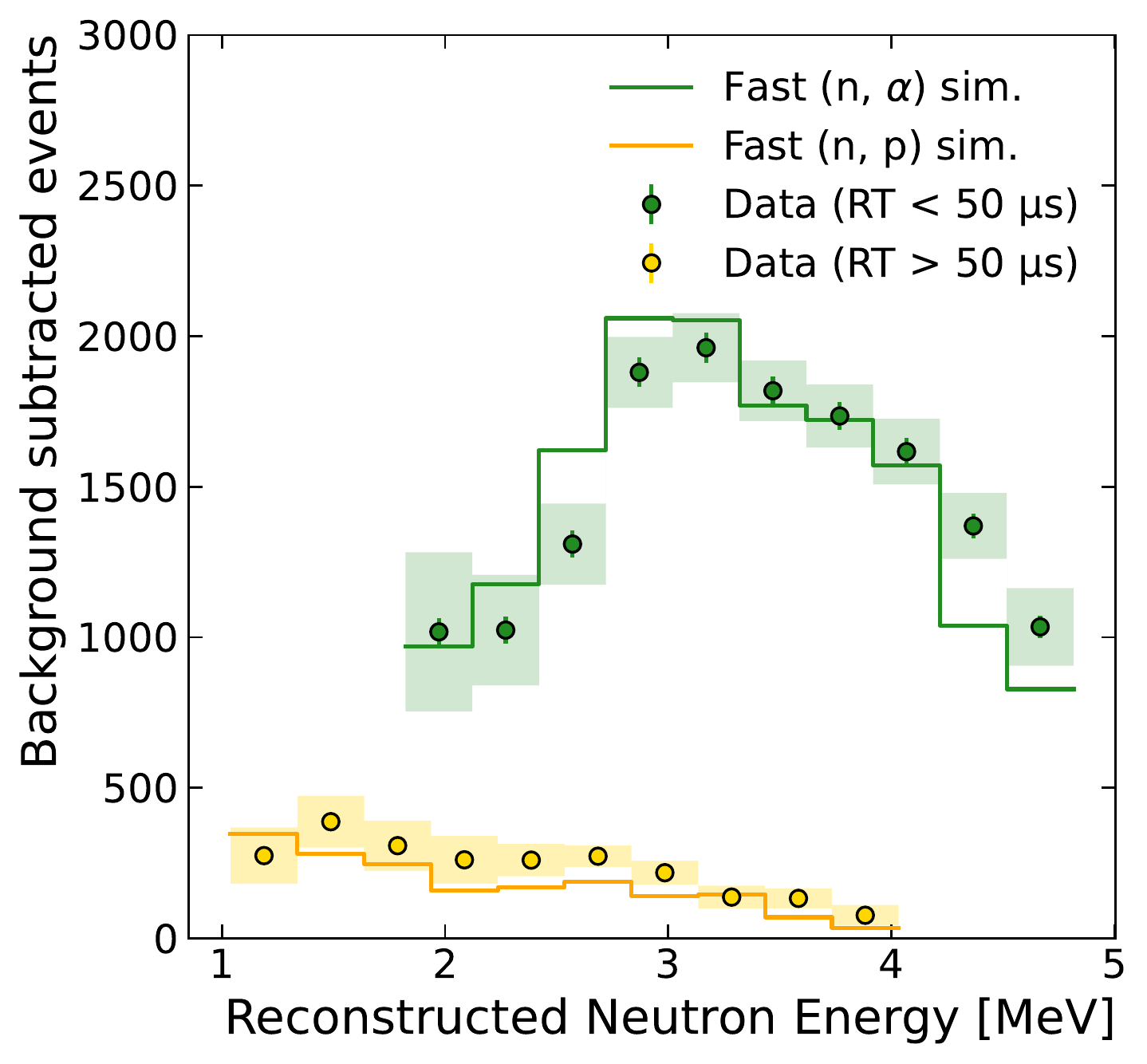}
\caption{The background subtracted data compared to the simulation prediction for the reconstructed neutron energy of the (n,p) and (n,$\alpha$) reactions. The rise time value is used to discriminate events attributed to each process.\label{fig:fig13}}
\end{figure}
Neutron capture in the vessel of the detector can produce $\gamma$-rays that interact in the gas volume, presented in purple. Nevertheless, these are rejected in the experimental data by the trigger threshold.

The portion of thermal neutrons that do not deposit all their energy in the volume of the detector, the aforementioned wall effect, was quantified with simulation studies at several pressure values. Results, presented in Figure \ref{fig:fig10b}, show how the impact of the wall effect decreases with the increase of the nitrogen pressure, with the fraction of events affected by the wall effect estimated to be 3.3\% at 1.8\,bar, compared with 5.7\% at 1.0\,bar and 10.4\% at 0.5\,bar, implying the reach of a plateau beyond approximately 2.5\,bar pressure for the specific 30\,cm diameter SPC. 

Fast neutron detection with the $^{241}$Am-$^{9}$Be neutron source and the spherical proportional counter filled with nitrogen at 1.5\,bar and 4.5\,kV anode voltage is demonstrated in Figure \ref{fig:fig11}. In this figure a comparison of the fast and thermalised neutron pulse amplitude distribution is presented. The excess  of events due to the detection of fast neutron is demonstrated with black solid line, compared to thermal neutron distribution with red dashed line. The response of the SPC to fast neutrons was investigated with the use of simulation, and the origin of the events is attributed to the corresponding reaction, as shown in Figure \ref{fig:fig12}. At the pulse amplitude distribution of Figure \ref{fig:fig12a}, the dominance of the (n,$\alpha$) reaction is evident for events with detected energy over 2\,MeV. 
The amplitude distribution of neutrons thermalised by the graphite stack (attributed to the (n,p) reaction) is obtained from simulation, while the corresponding  distribution of $^{222}$Rn decay chain events is obtained from a data taking run without the presence of fast neutrons from the source. Both of these contributions are normalised to the observed data. The dominant reaction at the first bin -denoted as ``Other''- represents $\gamma$-particles produced by neutron capture at the detector vessel and elastic scatterings in the volume of the gas. The rise time of the events with amplitude between the vertical dashed lines, that correspond to the detection of fast neutrons, is presented in Figure \ref{fig:fig12b}. In these figures, the (n,p) contribution is scaled by a factor 1.7, with respect to the cross section presented in Figure \ref{fig:fig2a}, to match the observed data. 
Nevertheless, it is clear that a threshold on approximately $50\,\si{\micro \textrm{s}}$  discriminates the (n,p) and (n,$\alpha$) processes, and enables a correction to be applied for the  $Q$-values of the corresponding reactions. As a result the energy of the incoming neutron is reconstructed and the observed spectrum is the convolution of the $^{241}$Am-$^{9}$Be spectrum with the cross section of the (n,p) and (n,$\alpha$) reactions respectively and the detector resolution. This is shown in Figure \ref{fig:fig13} for the events denoted as ``Fast'' in Figure \ref{fig:fig12a}, where the shadowed area represents a 50\% systematic uncertainty in the rates of the $^{222}$Rn and ``Other'' subtracted backgrounds. In the future the unfolding of the cross section and the improved detector resolution will help to obtain the underlying energy spectrum.

\begin{figure}[h]
\centering
\includegraphics[width=0.48\linewidth]{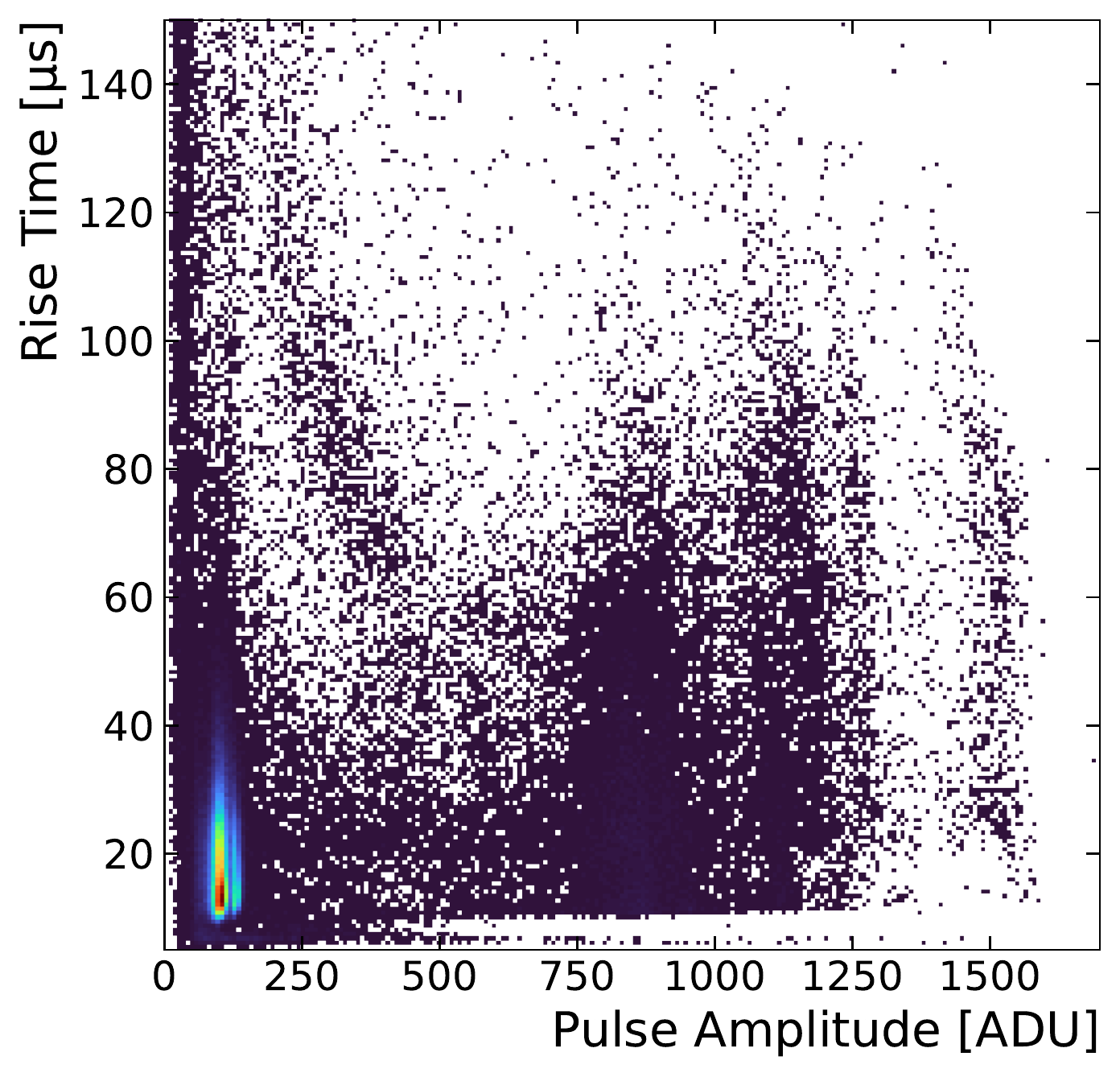}
\caption{Pulse rise time versus amplitude following the modified pulse processing, showing thermal neutrons events with 1.5\,bar N$_{2}$ and 4.5\,kV at the anode at approximately 110\,ADU. Additional to the thermal neutrons, the $^{222}$Rn decay chain is shown, ranging from 800 to 1600\,ADU.\label{fig:fig14}}
\end{figure}
 
\begin{figure}[h!]
\centering
\subfigure[\label{fig:fig15a}]{\includegraphics[width=14pc]{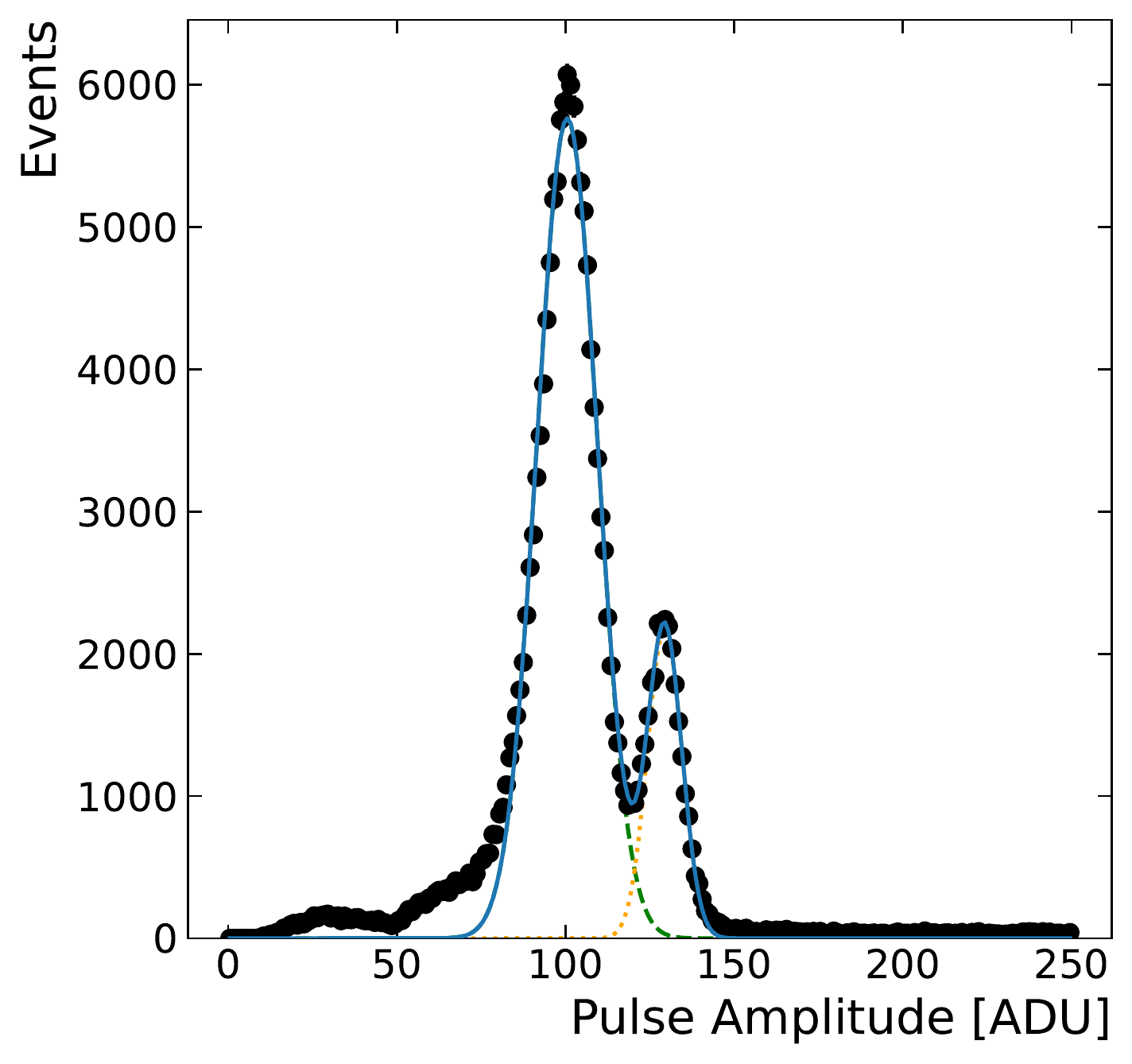}}
\subfigure[\label{fig:fig15b}]{\includegraphics[width=14pc]{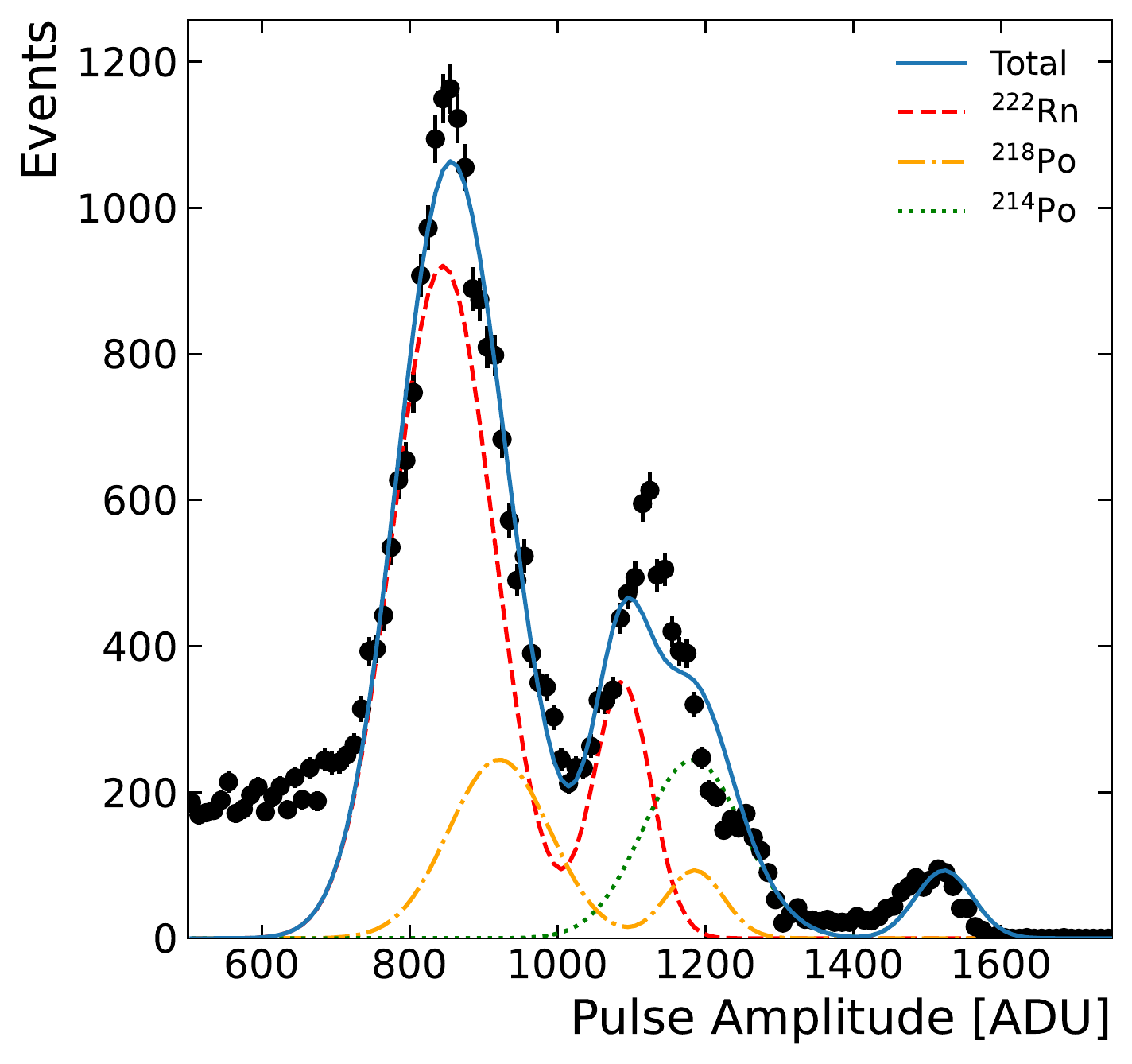}}
\caption{\subref{fig:fig15a} Thermal neutrons pulse amplitude distribution following the modified pulse processing at 1.5\,bar pressure and 4.5\,kV anode voltage.\subref{fig:fig15b} $^{222}$Rn decay chain peaks amplitude distribution. The double peak produced by thermal neutrons (Figure \subref{fig:fig15b}) is scaled to the $^{222}$Rn (dashed line), $^{218}$Po (dash-dotted line) and $^{214}$Po (dotted line) energies, while the sum of these distributions (solid line) agrees with the data.  
\label{fig:fig15}}
\end{figure}

Following the comparison between data and simulation, an investigation of pulses revealed a noise component. A modified  procedure for the estimation of the pulse shape parameters was derived, which implemented a low pass filter to remove high frequency noise and focused in the region around the pulse leading edge. The result of this approach is shown in Figure \ref{fig:fig14}, with the pulse rise time being in good agreement with the simulation. In Figure \ref{fig:fig15a} the thermal neutron peak is shown, which exhibit a much improved resolution.  Nevertheless, a second peak appears (with dotted line) which is attributed to the different response of one of the anodes. This is supported by the integral of the small peak being approximately the 1/4 of the integral of the main peak, that embodies the response of the 4 other anodes. This was already apparent from the  tail in the right side of the Gauss distribution in Figure \ref{fig:fig6b}, which was obscured by the deteriorated resolution. In Figure \ref{fig:fig15b} the $^{222}$Rn decay chain is presented with a fit model derived from the observation in Figure \ref{fig:fig15a}. This non uniformity in the response among anodes can be corrected for by individually reading out each anode and supplying an appropriate anode voltage. This has been also discussed in a recent publication~\cite{achinos} and will be our next R\&D goal. Nevertheless, in order to avoid obfuscating the main objective of this publication, that is the detection of neutrons at high pressure, this was not applied to the data.

\section{Conclusions and future efforts}
The spherical proportional counter operated with nitrogen is a promising alternative to $^3$He detectors for neutron spectroscopy, thanks to its simplicity, robustness and large volume. The measurement principle relies on the $^{14}$N(n,$\alpha$)$^{11}$B and $^{14}$N(n,p)$^{14}$C reactions. Earlier work has demonstrated the feasibility of the concept, but was limited to low pressure operation, due to the challenges in obtaining adequate signal amplification.
In this work, benefiting from the latest instrumentation developments, including the multi-anode sensor ACHINOS, the operating range for neutron spectroscopy was extended up to 1.8\,bar. 
Measurements were performed at the University of Birmingham, using an $^{241}$Am-$^{9}$Be source, and the obtained results were compared with dedicated simulations, presenting increased detection efficiency and mitigating the wall effect, capitalising on the increased target mass compared to $^3$He detectors. The (n,$\alpha$)~-~(n,p) discrimination capability  was demonstrated that enables the reconstruction of the energy of the incoming neutron. The improved  energy resolution anticipated with the use of individual anode read-out, will significantly improve the neutron spectroscopy performance of the system. This will be best demonstrated with the planned measurements using mono-energetic neutrons. The presented results are a leap towards the deployment of the spherical proportional counter as a neutron spectrometer at a variety of scientific, industrial, and medical settings. 

\section*{Acknowledgments}

This project has received funding from the European Union's Horizon 2020 research and innovation programme under the Marie Sk\l{}odowska-Curie grant agreements No 845168 (neutronSphere), No 841261 (DarkSphere) and No 101026519 (GaGARin).

\bibliography{mybibliography}

\end{document}